\documentclass[twocolumn]{aastex62}
\usepackage{lipsum}
\usepackage{graphicx}
\usepackage{float}
\usepackage[tight,TABTOPCAP]{subfigure}
\usepackage{fancyhdr}
\usepackage{hyperref}
\usepackage{enumitem}
\usepackage{amsmath}
\usepackage{amsfonts}
\usepackage{amssymb}
\usepackage{siunitx}
\usepackage{afterpage}
\usepackage{color}
\newcommand{\flash}{{\sf FLASH}}
\newcommand{\ramses}{{\sf RAMSES}}
\newcommand{\jina}{{\sf JINAbase}}

\begin{document}
\turnoffeditone

\title{\LARGE \textbf{The Effects of Metallicity and  Abundance Pattern of the ISM on Supernova  Feedback}}
\author{Platon I. Karpov}
\affiliation{Department of Astronomy $\&$ Astrophysics, University of California, Santa Cruz, CA 95064, USA}
\affiliation{Niels Bohr Institute, K\o benhavns Universitet, Copenhagen, Denmark}
\author{Davide Martizzi}
\affiliation{Department of Astronomy $\&$ Astrophysics, University of California, Santa Cruz, CA 95064, USA}
\affiliation{Niels Bohr Institute, K\o benhavns Universitet, Copenhagen, Denmark}
\author{Phillip Macias}
\affiliation{Department of Astronomy $\&$ Astrophysics, University of California, Santa Cruz, CA 95064, USA}
\affiliation{Niels Bohr Institute, K\o benhavns Universitet, Copenhagen, Denmark}
\author{Enrico Ramirez-Ruiz}
\affiliation{Department of Astronomy $\&$ Astrophysics, University of California, Santa Cruz, CA 95064, USA}
\affiliation{Niels Bohr Institute, K\o benhavns Universitet, Copenhagen, Denmark}
\author{Anne N. Kolborg}
\affiliation{Niels Bohr Institute, K\o benhavns Universitet, Copenhagen, Denmark}
\author{Jill P. Naiman}
\affiliation{Harvard-Smithsonian Center for Astrophysics, Cambridge, MA 02138, USA}
\begin{abstract}
Supernova (SN) feedback plays a vital role in the evolution of galaxies. While modern cosmological simulations capture the leading structures within galaxies, they struggle to provide sufficient resolution to study small-scale stellar feedback, such as the detailed evolution of SN remnants. It is thus common practice to assume subgrid models that are rarely extended to low metallicities, and which routinely use the standard solar abundance pattern. With the aid of 1-d hydrodynamical simulations, we extend these models to consider low metallicities and non-solar abundance patterns as derived from spectra of Milky Way stars. For that purpose, a simple, yet effective framework has been developed to generate non-solar abundance pattern cooling functions. We find that previous treatments markedly over-predict SN feedback at low metallicities and show that non-negligible changes in the evolution of SN remnants of up to $\approx 50\%$ in \textit{cooling mass} and $\approx 27\%$ in \textit{momentum injection from SN remnants} arise from non-solar abundance patterns. We use our simulations to quantify these results as a function of metallicity and abundance pattern variations and present analytic formulae to accurately describe the trends. These formulae have been designed to serve as subgrid models for SN feedback in cosmological hydrodynamical simulations.
\end{abstract}

\section{Introduction}
In order to study galaxy evolution, it is fundamental to understand the leading factor contributing to the observed structure of the interstellar medium (ISM): supernovae. These extreme events inject energies $E\approx 10^{51}\,{\rm erg}$ into the ISM, affect its phase structure \citep[e.g.,][]{McKee1977} and drive interstellar turbulence \citep[e.g.,][]{Joung2006,Faucher2013}. Consequently, SNe also have significant feedback on the star formation activity of galaxies \citep[e.g.,][]{Krumholz2005} and often drive galactic winds \citep[e.g.,][]{Veilleux2005}.

In the local universe, we have exquisite data on supernova remnants (SNRs) observed at multiple wavelengths \citep[e.g.,][]{Lopez2011,McCray2016,Lopez2018}. These results are generally interpreted within a framework that assumes relative abundance ratios of different chemical elements to be the same as in the Sun for any given metallicity \citep[e.g.,][]{MartizziSubGrid2015}. This is a reasonable assumption for stars that formed from well mixed gas, but it may not apply to low-metallicity environments, such as SNe from the earliest epochs \citep[e.g.][]{Shigeyama1998}. For example, metal-poor stars tend to have an enhanced abundance of $\alpha$-elements (e.g. C, N, O, Mg, Ca, Si), thanks to higher-mass stars' short lifetime enriching the environment through core-collapse events before low-mass stars can stabilize the abundance ratios to solar values through SN Ia explosions later in time. \citep[e.g.][]{Frebel2013}

To this end, we have two overall concerns. One is the validity of the most commonly used SN feedback prescription at low metallicities. The other one is the differences in feedback prescriptions that can arise from non-solar patterns

Feedback prescriptions  are constructed  because  it is significantly more tractable to assume a sub-grid model rather than accurately calculating SN feedback at the relevant physical scales. This is because a full calculation of the SNR evolution would be too expensive. Motivated by this,  we first plan to test in this paper the validity of the  low-metallicity sub-grid model extrapolations that are used today \citep[e.g.,][\ramses]{MartizziSubGrid2015,Martizzi2016}.


Next, in contemporary astrophysics numerical simulations are often used to study the formation and evolution of galaxies, as well as their gas, stars and chemistry. Currently, it is fairly standard for the community involved in cosmological simulations to assume solar composition, and scale it to a particular degree, in regards to evaluating the chemical yield of SNe, e.g. {\sf Illustris} \citep{Vogelsberger2013, Vogelsberger2014}, {\sf MUFASA} \citep{MUFASA2016} and {\sf Horizon-AGN} \citep{Dubois2014}. Additionally, solar element abundance patterns are typically assumed when computing metal-dependent radiative cooling, which influences the dynamics and cooling rates of SNRs. That being said, it is unlikely for the ISM of primordial galaxies to have started with solar abundance ratios \citep[e.g.,][]{2015ApJ...807..115S,2018MNRAS.477.1206N}. The solar abundance pattern is the result of multiple generations of stars that have been contributing to and mixing within the ISM, enriching it closer to solar abundance ratios. Thus, significant differences in metallicity and abundance patterns are to be expected for different generations of stars \citep[e.g.,][]{Komiya2011}. This information is important when trying to constrain the chemical composition of the ISM where these stars formed, which was polluted by the SNRs triggered by a previous population of stars \citep[e.g.,][]{Kobayashi2006}.

Since gas cooling is metal-dependent, and since the evolution of SNRs is determined by cooling of the ISM swept by the SN ejecta generated shock, both factors have to be taken into account. In this paper, we choose to study in isolation the effects of solar and non-solar ISM abundance patterns by modeling cooling functions for arbitrary mixtures of gas, and then we plug this machinery into numerical simulations of SNRs. This approach allows us to model SNRs in a wide range of environments in terms of metallicities and abundance patterns.

To summarize our goals, in this paper we set out to:

\begin{itemize}
    \item build a new model for the dynamics of SNRs that is more accurate at low metallicity than available methods, assuming a solar-abundance pattern (Section \ref{sec:solar})
    \item explore the low-metallicity regime with the newly derived model, checking the SNR behavior in a range of temperature environments  (Section \ref{sec:ejecta})
    \item study the effects of the non-solar abundance pattern for SNR evolution at all metallicities (Section \ref{sec:nonsolar})
\end{itemize}

The paper describes our method of generating cooling functions with arbitrary composition, along with the hydro setup within \flash\ in Section \ref{sec:methods}. The Results, as outlined by our goal summary above, can be found in Section \ref{sec:results}. Discussion in Section \ref{sec:discussion} describes the implication of these results onto galaxy star formation simulations, tested in \ramses\, and the implications of a non-solar abundance pattern onto SN feedback. Section \ref{sec:conclusion} summarizes our conclusions.

\section{Methods}
\label{sec:methods}
\subsection{Cooling Function Calculation}
It is an integral portion of our setup to accurately model energy loss due to radiative cooling. This work samples a variety of chemical compositions of the ejecta and ISM, projecting their evolutionary tracks in regards to them mixing and cooling for the next generation of stars to form. Therefore, a convenient python tool has been developed that uses the latest atomic data.
\begin{figure}
\centering
\includegraphics[width=0.9\linewidth]{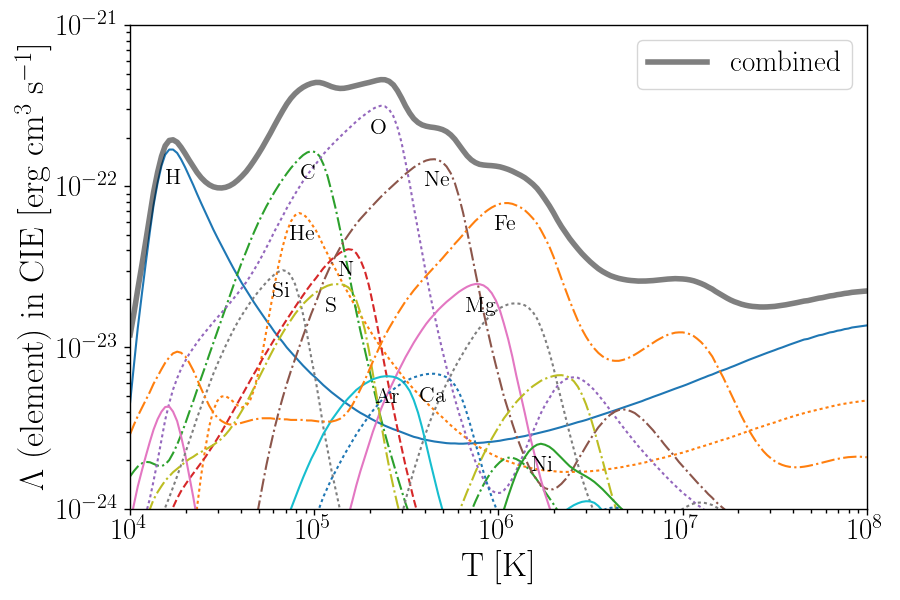}
\caption{Element-by-element cooling assuming CIE and solar elemental abundance ratios. The total CIE cooling efficiency due to all elements is shown by the upper thick gray curve.}
\label{fig:regnat}
\end{figure}

Consider ions of a particular element.  Let $n_{\rm ion}$ be the number density of ions with ionization state $j$.  We are interested in tracking the change of $n_{\rm ion}$ with temperature.  Collisional ionization will result in promoting ion's state from $j$ to $j+ 1$, while recombination does the opposite. Our setup assumes a system in  collisional ionization equilibrium (CIE). The general form is summarized by:
\begin{eqnarray}
n_{\rm e} n_{\rm ion}^{j+1} \alpha_{\rm ion, j}(T)=\xi_{\rm ion, j}n_{\rm ion}^j+n_{\rm e} n_{\rm ion}^j C_{\rm ion, j}(T)
\end{eqnarray}
where $n_e$ is the number density of electrons, $\alpha$ is a recombination coefficient, $\xi$ is a photoionization coefficient, and $C$ is the collisional ionization coefficient. For our case, we only consider recombination and collisional ionization processes, assuming no nearby sources to effectively photoionize surrounding gas beyond the temperature floor of $10^4\,{\rm K}$. Thus, CIE can be written as
\begin{eqnarray}
n_{\rm e} n_{\rm ion}^{j+1}\alpha_{\rm ion,j}(T)=n_{\rm e} n_{\rm ion}^jC_{\rm ion,j}(T).
\end{eqnarray}

Since the system is in CIE, relative temperature-dependent abundance fractions of the ions for each element will remain constant. Thus, CIE eliminates the need to calculate the ion abundance ratios at every timestep, allowing us to use the ionization fraction tables from \citet{Bryans2009}. Furthermore, we used the most up-to-date ion-by-ion cooling function tables from \citet{Gnat2012} calculated by Cloudy, which account for collisional excitation and line emission, ion recombination, collisional ionization, and Bremsstrahlung radiation. This allows us to calculate the total cooling function efficiency for any composition in CIE.

We have calculated the cooling efficiencies for each ion as a function of temperature, where  the total amount of energy lost  per ion due to radiative cooling typically takes the following form
\begin{eqnarray}
\left(\frac{dE}{dVdt}\right)_{\rm ion} = n_{\rm e} n_{\rm ion} \Lambda_{{\rm e, ion}}(T,Z/Z_{\odot}),
\end{eqnarray}
where on the left hand-side there is energy [erg] per unit volume [cm$^3$] per unit time [s]; on the right hand side $n_e$ is the free electron number density [cm$^{-3}$], $n_{\rm ion}$ is the number density of  an ion [cm$^{-3}$], and $\Lambda$ is the cooling function itself [erg cm$^3$ s$^{-1}$], which is dependent on temperature [K], metallicity [Z], and the abundance pattern.

This formalism can be used to compute the total  cooling efficiency for any arbitrary composition \citep{Gnat2012}. For example, Fig.~\ref{fig:regnat} shows the ion-by-ion CIE cooling efficiencies, $\Lambda_{e,{\rm ion}}$, due to different elements under the assumption of solar metallicity \citep{Gnat2012}. It is important to note that we are considering the temperature range above $10^4\,{\rm \edit1{K}}$ since the temperature floor of the ISM is assumed to be kept at $10^4\,{\rm K}$ due to combined photoionization from the surrounding stars. In addition, Table \ref{tb:hierarchy} lists the elements in the order of their contribution to the overall cooling function under the assumption of solar metallicity.

\begin{table}
\centering
\begin{tabular}{c c c}
\hline
\hline
Rank	&	Element	&	Contribution	\\
\hline
1	&	H	&	\num{4.3e-1}	\\
2	&	Fe	&	\num{1.8e-1}	\\
3	&	He	&	\num{1.6e-1}	\\
4	&	O	&	\num{6.2e-2}	\\
5	&	Ne	&	\num{6.0e-2}	\\
6	&	Si	&	\num{2.9e-2}	\\
7	&	S	&	\num{2.0e-2}	\\
8	&	Mg	&	\num{1.9e-2}	\\
9	&	C	&	\num{1.3e-2}	\\
10	&	Ni	&	\num{9.7e-3}	\\
11	&	Ca	&	\num{6.9e-3}	\\
12	&	N	&	\num{5.2e-3}	\\
\hline
\hline
\end{tabular}
\caption{Hierarchy of elements according to their fractional contribution to the overall solar abundance cooling function. The values come directly from Fig.~\ref{fig:regnat}}.
\label{tb:hierarchy}
\end{table}

In order to simplify cooling calculations within a hydrodynamical code framework, our cooling curve calculation outputs $\lambda(T)$, which is the total energy lost per unit density squared.

\begin{eqnarray}
\frac{dE}{dVdt} = \rho^2 \lambda(T)
\end{eqnarray}
The advantage of having this form is that we can calculate in advance the total normalized cooling of the system. All that needs to be done within the main simulation is to multiply $\lambda$ by the readily available $\rho^2$ of the cell. The setup to generate any composition cooling function curve in CIE can be found on github\footnote{\url{https://github.com/pikarpov/CoolingCurve}}.

\subsection{Hydro Setup}
For this work we make use of the \flash\ hydrodynamical code \citep{Fryxell2000} in order to calculate the  1-d  cooling evolution of an expanding supernova remnant, starting at the Sedov phase and ending well after the radiative phase.

\subsubsection{Technical details}
We are using a 1-d spherical adaptive mesh refinement (AMR) grid with a maximum of 8 levels of refinement, and a maximum of $10^3$ cells, which proved to be a sufficiently high resolution for our study. The solver is the 5$^{th}$ order piecewise parabolic method (PPM), with a Harten-Lax-van Leer-Contact (HLLC) Riemann solver.

Considering the absence of a CIE cooling module within \flash\, we have developed it from scratch and have made it publicly available. The current implementation tracks the mixing of two species: expanding ejecta and the ISM. This setup allows us to examine how new metals from the ejecta are enhancing the metal fraction within the ISM that resulted from the enrichment of previous generations of stars. Thus, given initial abundances for the species, the code calculates the appropriate cooling functions to use within the simulation. Based on species' mass fractions in each cell, effective cooling is then calculated and subtracted at each time-step, as the energy is being radiated away. As commonly implemented, an additional timestep condition limiting energy loss is added in order to prevent the cells from radiating a sizable fraction of their internal energy.

\subsubsection{Homogeneous ISM setup}
\label{sec:setup}
The parameters were chosen in order to conduct a detailed comparison with the work of \citet{MartizziSubGrid2015}. Thus, we injected $M_{\rm ej}=3\,M_\odot$ with thermal energy of $E_{\rm th}=10^{51} \, {\rm erg}$ within a spherical region of $10\;\Delta x$, where $\Delta x$ constitutes the radial extent of a cell. Even though there is no kinetic energy being injected, the Sedov-Taylor profile is quickly reached with proper $E_{\rm th}$ and $E_{\rm kin}$ ratios established. Empirical evidence from our simulations showed that it takes for ejecta to sweep $\approx 5$ times its initial mass to reach Sedov stage. After the proper density, pressure, and velocity self-similar profiles have been established, the cooling is turned on. By default, we set ISM to contain 10$^2$ particles per cm$^3$, but a large range of densities was systematically  explored in order to derive the fitting formulas presented in Section \ref{sec:fitting_formulae}.

Our simulations follow through the SNR evolution stages that can be broadly characterized with a power-law relationship between the shock radius and time, $R_s\propto t^{\eta}$ \citep{cox1972, Cioffi1988}:
\begin{enumerate}[itemsep=0mm]
\item \textbf{Free expansion}: the inertia of the SN ejecta dominates the expansion, with its mass being greater than the swept up mass ($\eta=1$)
\item \textbf{Sedov-Taylor}: self-similar profiles of pressure, density and velocity are established and due to negligible cooling, the total energy content remians constant ($\eta=\frac{2}{5}$)
\item \textbf{Radiative phase}: shock pressure driven expansion, during which cooling becomes important  ($\eta=\frac{2}{7}$) \label{enum:snowplow}
\item \textbf{Snowplow phase}: once a sufficient amount of energy has been radiated, total momentum (feedback) of the SNR converges to a constant value, while sweeping of the ISM material continues ($\eta=\frac{1}{4}$)
\end{enumerate}

The radius of the SNR at the time when cooling becomes efficient (stage \ref{enum:snowplow}), is the cooling radius, $R_{\rm cool}$, and the enclosed mass is referred to as the cooling mass, $M_{\rm cool}$. This allows for the calculation of feedback (momentum deposition) of a particular SNR. These are the primary variables that will be analyzed throughout the paper.

Our goal is to examine SN feedback at different regimes of ejecta and ISM abundance patterns, as well as densities and temperatures. The deliverable is a suite of fits to estimate feedback given the aforementioned parameters.

\section{Results}
\label{sec:results}
\subsection{Feedback using Solar Abundance Pattern}
\label{sec:solar}

\begin{figure}
\centering
\includegraphics[width=1\linewidth]{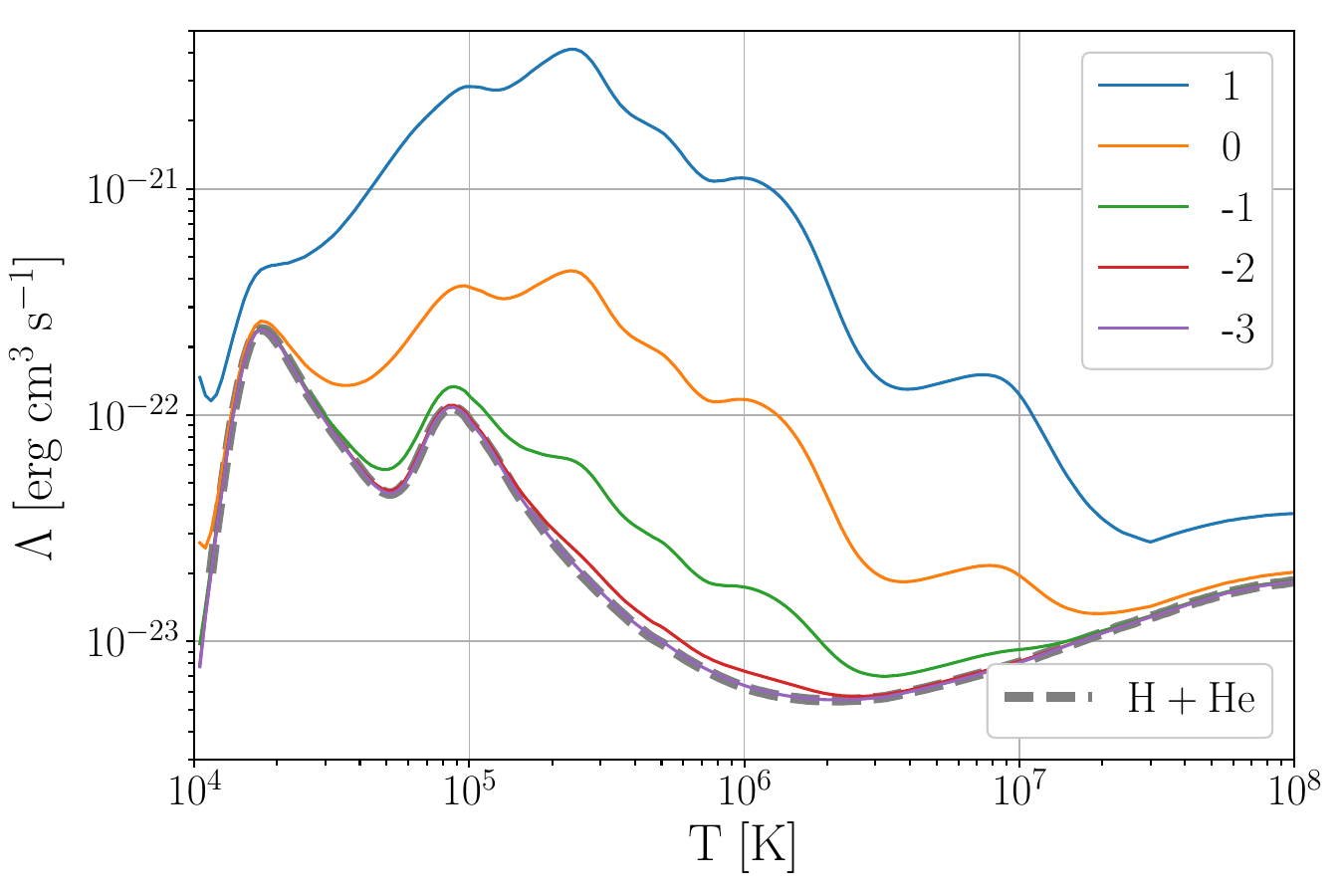}
\caption{Cooling functions for different solar abundance ratios. That is, ${\rm log}(Z/Z_{\odot})=[1,0,-1,-2,-3]$. When  $Z/Z_\odot$ $\lesssim 10^{-2}$, the cooling function quickly converges to the metal-free one.}
\label{fig:cool_frac}
\end{figure}

As it is typical for the community to assume a scaled solar abundance pattern within simulations of various epochs, we test our setup under this assumption. Fig.~\ref{fig:cool_frac} showcases \edit1{the cooling functions for different metal fractions}. As can be seen, at $Z/Z_\odot=10^{-3}$, the cooling function converges to the metal-free setup. Thus, in a metal-poor environment, H and He \edit1{will dominate} the cooling evolution of the remnant.

\begin{figure*}
\centering
\subfigure[Thermal Energy]{\includegraphics[width=1\columnwidth]{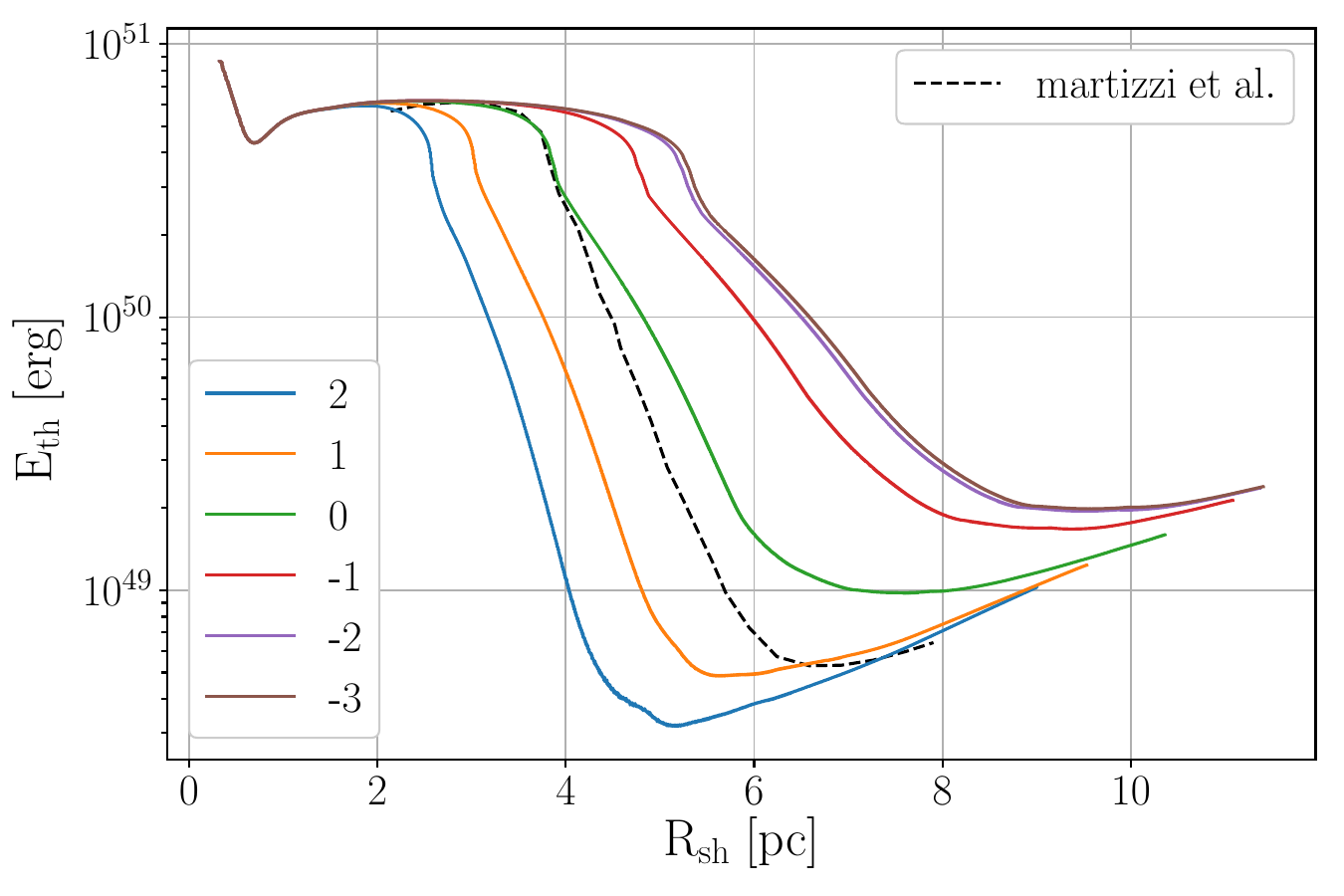}
\label{fig:z_eth}}%
\subfigure[Momentum]{\includegraphics[width=1\columnwidth]{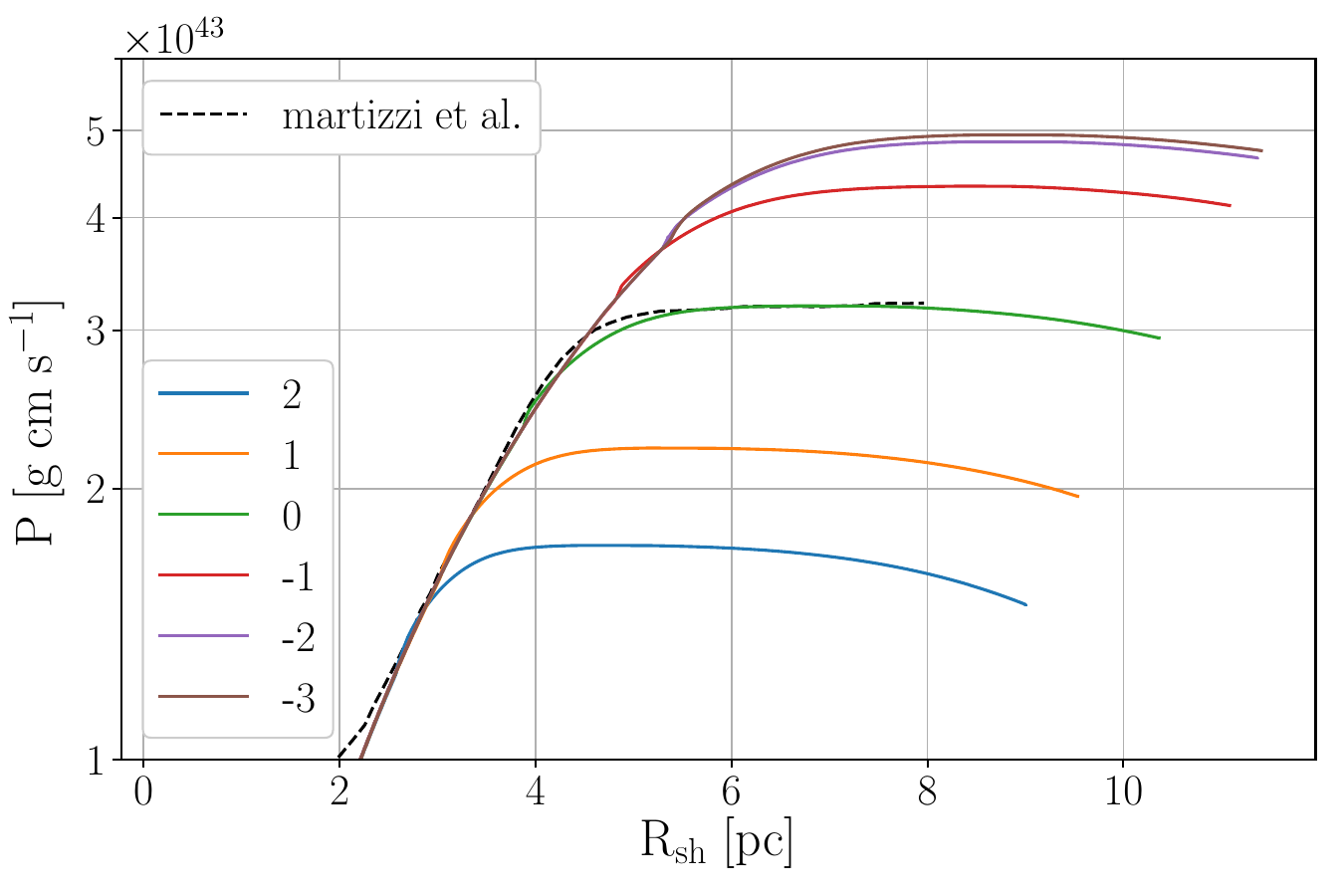}
\label{fig:z_momentum}}%
\caption{Thermal energy (left) and momentum (right) as a function of shock position, while varying log$(Z/Z_{\odot})$. Ejecta and ISM both have $Z=Z_{\odot}$.}
\label{fig:z_q}
\end{figure*}

Using these cooling functions, we evolve our 1-d hydro simulations to observe the cooling of the SNR. By looking into thermal energy ($E_{\rm th}$) and momentum ($P$) evolution versus shock radius ($R_{\rm sh}$), we can track the cooling efficiency and estimate stellar feedback for each simulation run (Fig.~\ref{fig:z_q}). The SNe are initialized with thermal energy $E_{\rm th}=10^{51}\,{\rm erg}$ and $E_{\rm kin}=0\,{\rm erg}$, but the system quickly relaxes to the Sedov-Taylor solution, in which total energy is conserved and thermal energy plateaus to $E_{\rm th}\approx 7.1\times 10^{50}\,{\rm erg}$, as also seen by \citet{MartizziSubGrid2015}. This comes from the shockwave conditions keeping the ratio of $E_{\rm th}$ and $E_{\rm kin}$ constant: $\frac{E_{\rm th}}{E_{\rm kin}}=\frac{4}{\gamma^2-1}$, that gives the energy split of $\approx 70\%$ $E_{\rm th}$ and $\approx 30\%$ $E_{\rm kin}$ for $\gamma=\frac{5}{3}$. In Fig.~\ref{fig:z_eth}, the sharp decay of $E_{\rm th}$ signifies the beginning of the regime in which cooling becomes efficient, hence marking the position of $R_{\rm cool}$. In our simulation, we quantify $R_{\rm cool}$ by taking it as \edit1{the} position of the shock once $\approx 30\%$ of $E_{\rm th}$ has been lost \citep{MartizziSubGrid2015}. The thermal energy profiles start \edit1{to} increase after reaching \edit1{global minima} in $E_{\rm th}$ since the rate of thermal energy increase from sweeping external medium is larger than the cooling rate. In Fig.~\ref{fig:z_momentum}, momentum deposition, i.e. feedback, is taken to be the maximum value of $P$. Fig.~\ref{fig:z_q} show that SNR evolution follows the trend of cooling functions converging at and below $Z/Z_\odot=10^{-3}$. In addition, \cite{MartizziSubGrid2015} 3D results are presented in both plots of Fig.~\ref{fig:z_q} to further validate our 1D study. \edit1{The initial conditions in \cite{MartizziSubGrid2015}} match ours from Section~\ref{sec:setup}, except for dimensionality, resolution, and the use of \citet{SutherlandDopita1993}  cooling functions instead of \cite{Gnat2012}. Please see Appendix~\ref{sec:res} for more details on the 1D vs. 3D comparison, where we also address the slope difference between the models at log($Z/Z_{\odot}$)=0.

\begin{figure*}
\centering
\subfigure[Cooling mass]{\includegraphics[width=1\columnwidth]{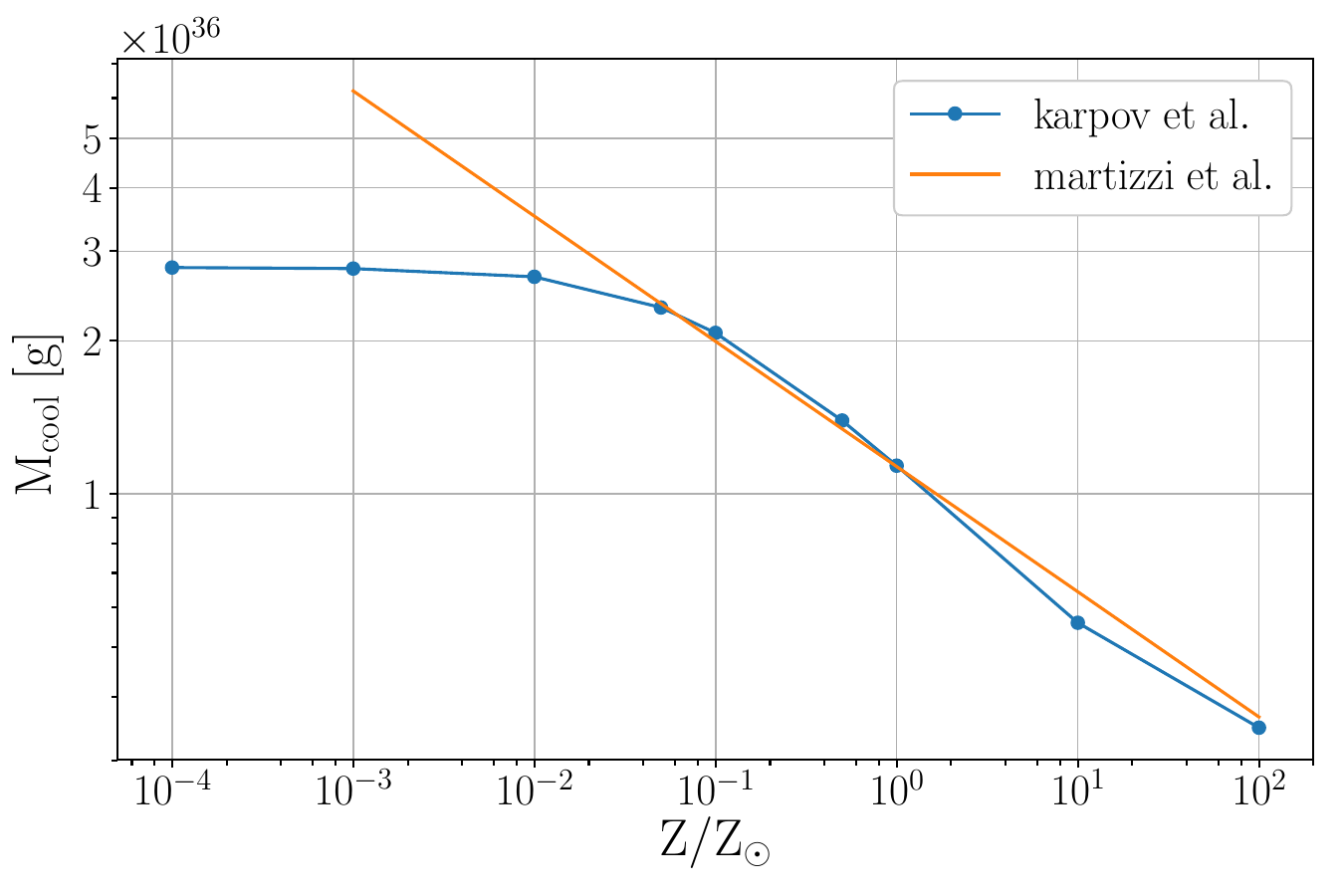}
\label{fig:Mcool_solar}}%
\subfigure[Momentum]{\includegraphics[width=1\columnwidth]{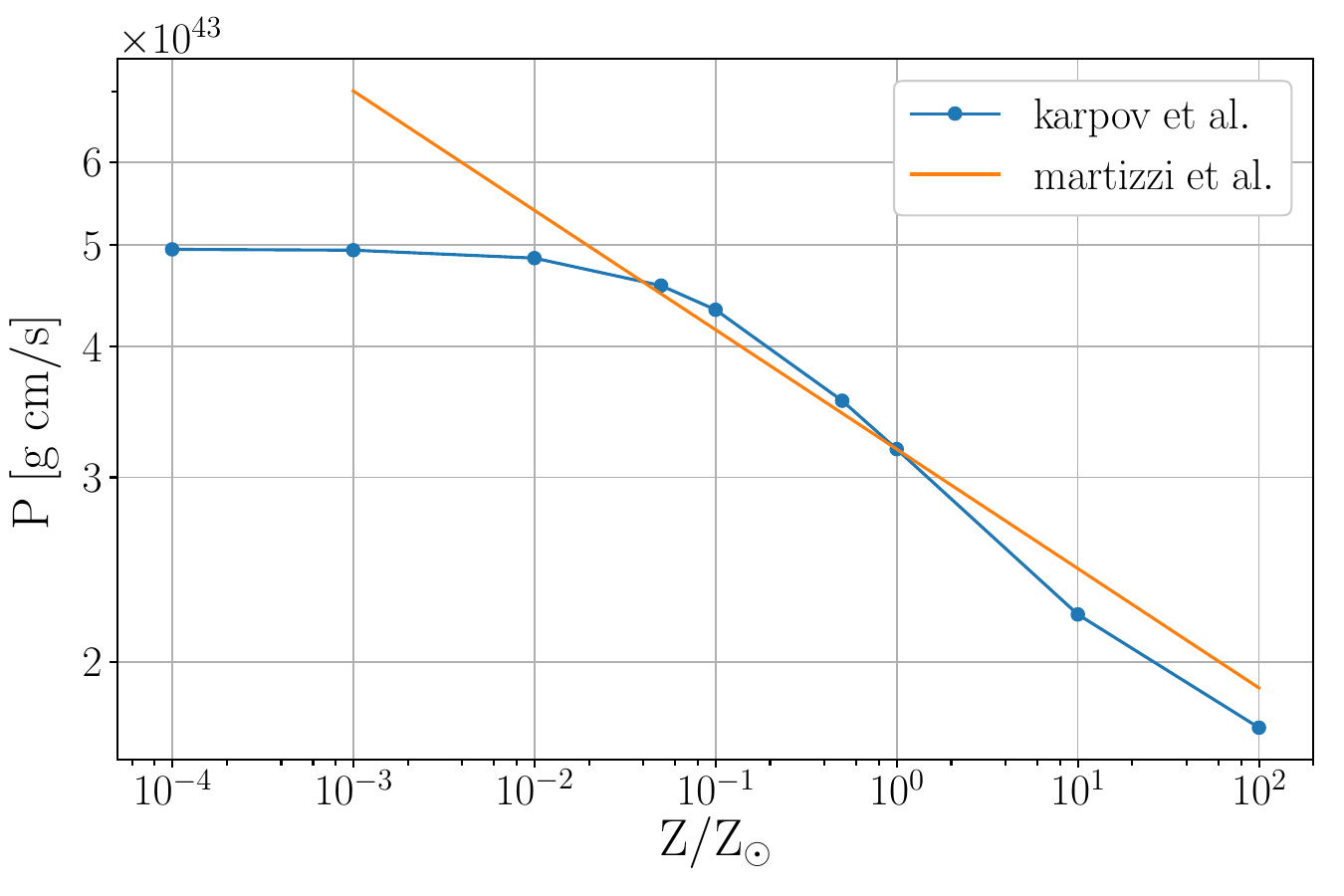}
\label{fig:Momentum_solar}}%
\caption{The converged cooling mass and momentum as a function of the solar abundance fraction, along with the commonly used power-law fit provided by  \cite{MartizziSubGrid2015} for each quantity. Ejecta and ISM both have $Z=Z_{\odot}$.}
\label{fig:q_solar}
\end{figure*}

Extracting $M_{\rm cool}$ and converged $P$ values from our simulations, a non-power-law trend can be observed (Fig.~\ref{fig:q_solar}). In those figures, the power-law fits from \cite{MartizziSubGrid2015} are overplotted, which were obtained by fitting the results of 3D numerical hydrodynamical simulations of isolated SNRs that fully resolve the Sedov-Taylor, radiative and snowplow phase of their evolution. The results of \cite{MartizziSubGrid2015} are in accordance with similar 1D simulations by \cite{Thornton1998}, but the authors only considered SNR evolving in the ISM with metallicity $Z\geq 0.1\,Z_{\odot}$, and they mention that their power-law fits might break down at lower metallicities. While the results from our 1D simulations are in good agreement with the power-laws of \cite{MartizziSubGrid2015} for environments with near solar metallicities, the discrepancies become significant for metal poor environments. Most current simulations of galaxy evolution use similar but somewhat less accurate SN momentum-driven models: \ramses\ \citep{Kimm2014}, {\sf FIRE} \citep{Hopkins2014}, {\sf Enzo} \citep{Simpson2015}, {\sf ART} \citep{Semenov2017}. This raises concerns with the need to evaluate potential errors in the feedback estimate within metal-poor environments. We turn our attention to this issue in Section \ref{sec:galev}

\subsection{The Cooling Contribution of Ejecta Metals}
\label{sec:ejecta}

\begin{figure*}
\centering
\hspace{-1.1 mm}
\begin{minipage}{.47\textwidth}
  \centering
\includegraphics[width=1\linewidth]{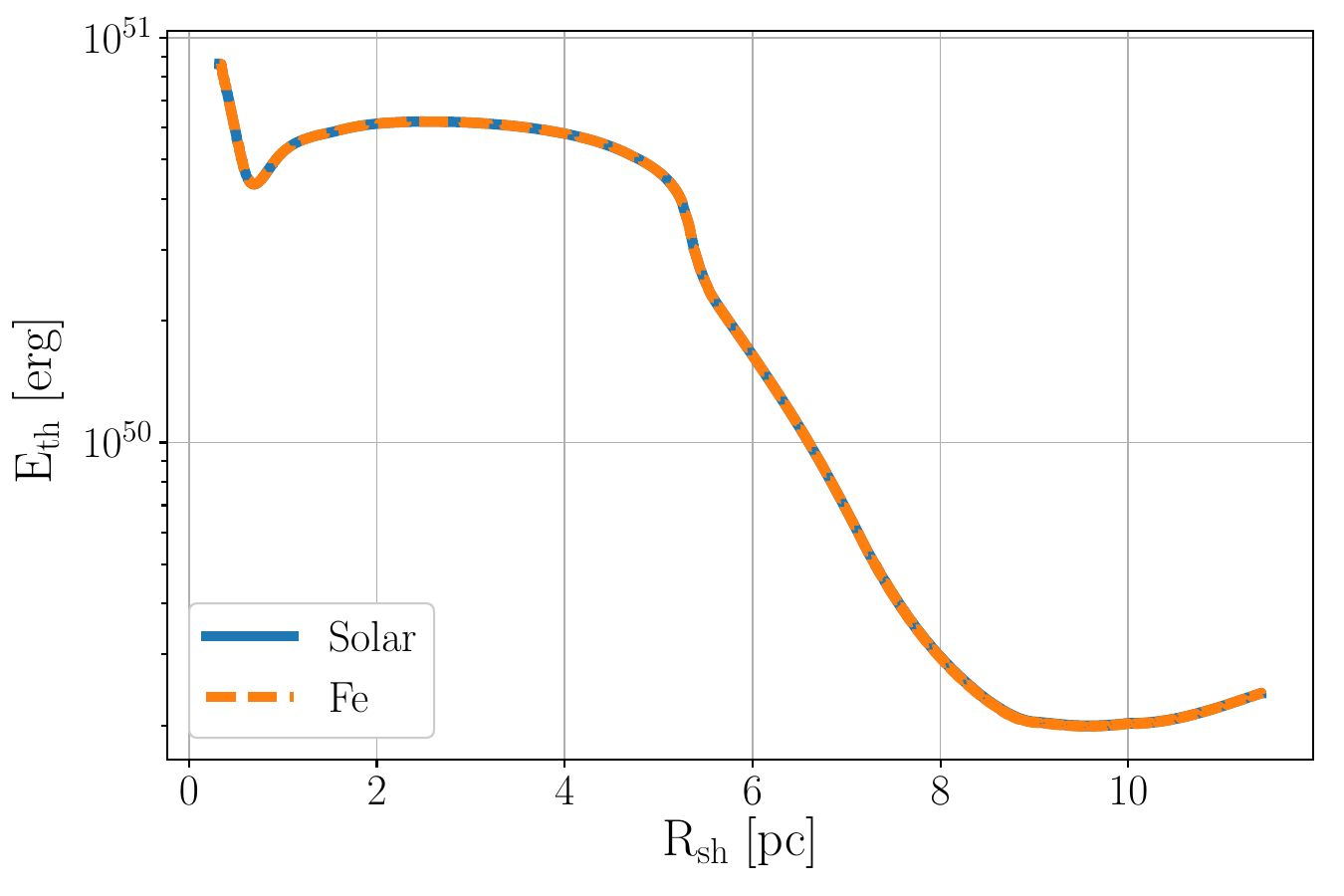}
\caption{Solid: Solar abundance ejecta; Dashed: pure Fe ejecta. Both expand into pristine H+He ISM. Since they lie on top of each other, the cooling contribution from the metals of the ejecta is negligible and most of the energy losses arise from the swept up ISM material.}
\label{fig:Fe_pristine}
\end{minipage}%
\hspace{0.045\textwidth}
\begin{minipage}{.47\textwidth}
  \centering
\includegraphics[width=1\linewidth]{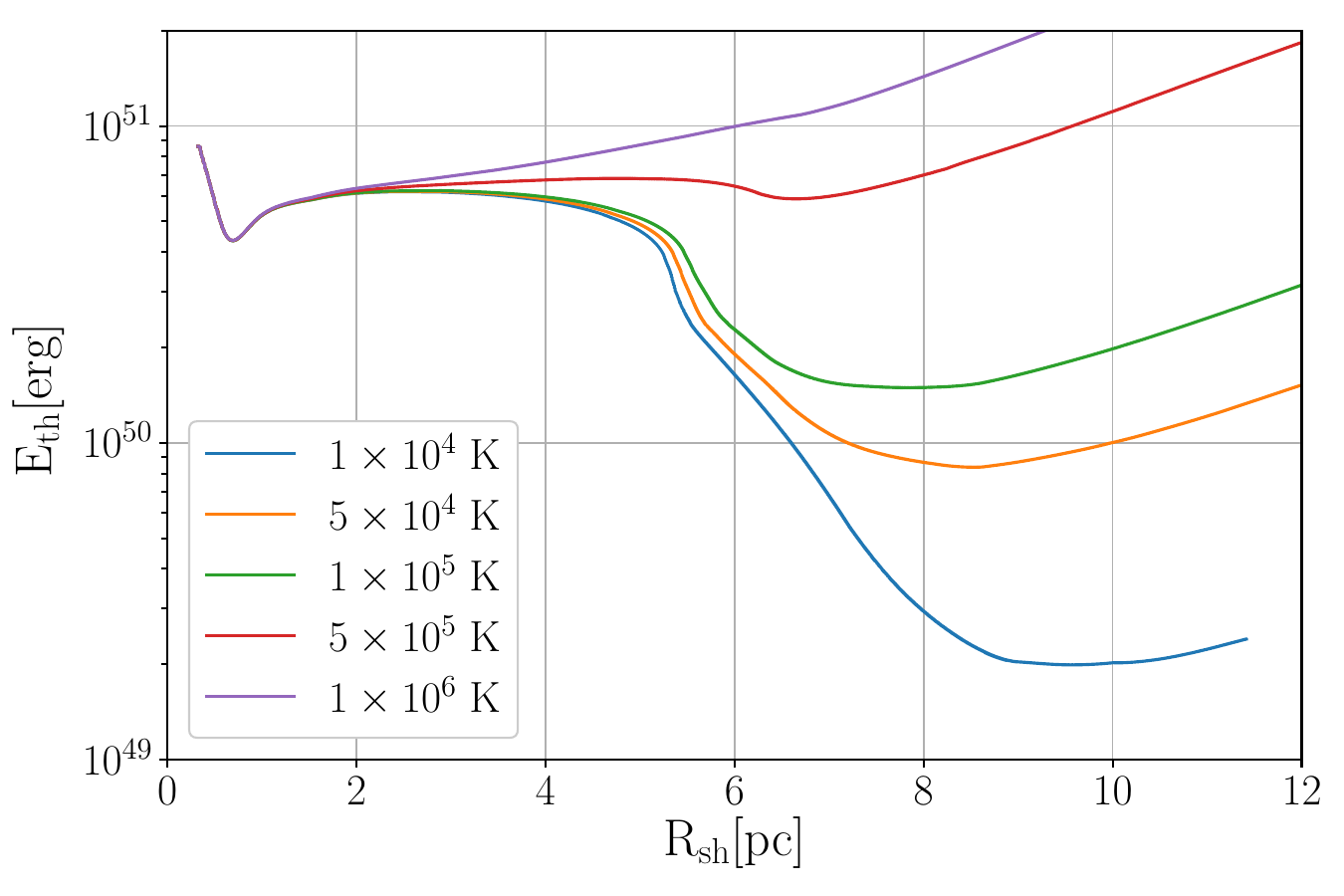}
\caption{Temperature dependence of the SNR expanding into pristine H+He ISM, tracking the changes in $E_{\rm th}$. Each curve was calculated at different values of the ambient T as denoted in the labels. Even in an extreme case, SNR evolution is dictated by the ISM.}
\label{fig:Tdep}
\end{minipage}
\end{figure*}

Considering Pop III stars, which explode in pristine (H+He only) environments \citep{annurev-astro-082812-140956}, it could be speculated that the metals from their ejecta might become a considerable factor in the SNR evolution. We have tested an extreme setup of pure Fe ejecta expanding into a pristine H+He ISM to test the effects of ejecta contribution to SN cooling. As can be clearly  seen in Fig.~\ref{fig:Fe_pristine}, the SNR evolution is independent of the ejecta composition. Thus, it is the ISM composition which completely determines the effectiveness of SN feedback, with the metals from the ejecta not making a significant contribution. However, cooling efficiency is highly dependent on temperature, hence effects of the ambient temperature of the surrounding medium need to be tested.

The calculations presented in Fig.~\ref{fig:Fe_pristine} assumed a temperature floor of $10^4$ K due to the assumption  of effective photoionization from the surrounding stars. However, there are instances where supernovae occur in much hotter environments. For example, consider an exploding older stellar population (Type Ia SNe) in a galactic halo, where the ISM temperature can be on the order of $10^6$ K \citep[e.g.,][]{Dorfi1996,Tang2005}. By looking at Fig.~\ref{fig:regnat}, it can be seen that at such high temperature, H and He cooling efficiency peaks are no longer contributing to the overall cooling function, due to the medium being too hot for these elements to recombine. In this setup we have tested the extreme scenario of a pure Fe ejecta exploding into a pristine H+He ISM with T$_{\rm ISM}$=: $10^4$, $5\times10^4$, $10^5$, $5\times10^5$ and $10^6\,\rm K$.

With increased temperature, the contribution of H and He to the overall cooling starts to diminish in comparison to metals, as shown in Fig.~\ref{fig:Tdep}. This is due to the maximum recombination rates (hence $\Lambda$ peak) of H and He occurring at lower temperatures, below $2\times 10^4\,\rm K$ and $10^5\,\rm K$ respectively, as can be seen in Fig.~\ref{fig:regnat}. However, cooling from the Fe ($\Lambda$ peaking around T=$10^6\,\rm K$) is modest and not able to effectively counterbalance the additional thermal energy  added by the newly swept-up material. We thus conclude that in most circumstances, cooling from the metals of the SN ejecta is a secondary contribution even in the most pristine environments, wheres the evolution of SNRs is dominated by cooling of the swept-up ISM.

\subsection{Feedback using Non-Solar Abundance Patterns}
\label{sec:nonsolar}

Following the metal enrichment of the Universe, it has been saturated to a solar abundance pattern rather recently relative to its inception. In the early times, strong deviations from a solar abundance pattern have been observed in metal-poor stars \citep{Sneden2008}. Hence, if one has interest in studying Pop III or even low-$Z$ Pop II stars, scaling solar abundance in accordance to the total metallicity has potential to produce significant uncertainties in feedback calculations, directly affecting galactic evolution timescales in the simulations.

For our test case, we have used a metal-poor stars' database called \jina\ \citep{JINAbase2018}. From there we assume the metal-poor stellar abundances to be tracers of the primordial ISM abundances. This will be used to constrain the effect of non-solar ISM chemistry on SN feedback and to place constraints on the chemistry of early galaxies. The method proposed here will produce valid results provided that the observed metal abundances  are those from the chemically primitive gas clouds from which the stars formed.  Using  metal-poor stars as tracer particles of the ISM could produce errors due to extrinsic mechanisms, such as  stellar convection bringing newly produced elements to the surface or binary mass transfer, thus changing a star's metallicity in comparison to that of the ISM from which it was formed \citep{Herwig2005,Placco2014}.  However, in such instances,  the results  of this study will provide  a robust  \textit{upper limit} on the effects of non-solar abundances patters on SN feedback.

\begin{figure}
\centering
\includegraphics[width=1.0\linewidth]{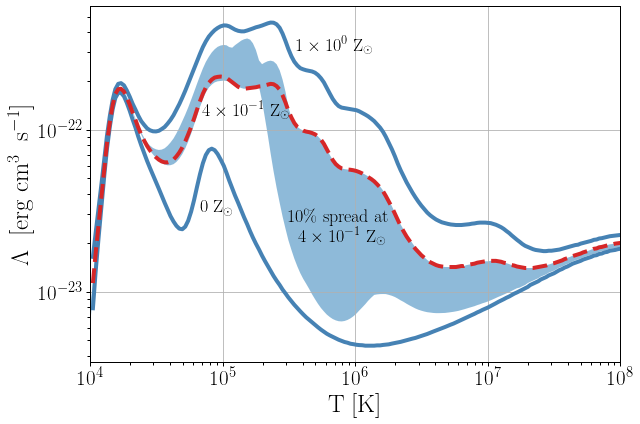}
\caption{
The plot stands to show an example of how differences in relative metal abundances can affect the overall cooling function near a set metallicity. Red-dashed line is the cooling function at $Z=4\times 10^{-1} \, Z_\sun$, assuming solar-abundance pattern. The blue shaded region shows the variations in the cooling function based on star samples from \jina, with metallicities $Z=4\times 10^{-1}\, Z_\sun\pm10\%$. For comparison we also plot the cooling functions for solar metallicity and pristine ISM.}
\label{fig:non_solar}
\end{figure}

As a result, we have generated a suite of SNR models with cooling functions, wildly varying in shape, from the near-pristine metal-poor regime to metal-enhanced, super-solar abundances. For the elements that were missing in the database, we assumed a solar pattern. A selection of non-solar abundance cooling functions is presented in Fig.~\ref{fig:non_solar} for a relative metallicty of $Z=4\times 10^{-1}\, Z_\sun$, chosen for the purpose of illustrating the potential spread in energy losses. That being said, we could have chosen any other metallicity for this purpose.

Since the Universe has not been enriched according to a solar abundance pattern, certain elements play a greater role in cooling the ISM than Table \ref{tb:hierarchy} might suggest. For example, while Mg has been enriched following a solar abundance pattern, the situation for C is vastly different, with its super-solar abundance dominating the cooling curves. In total, the elements responsible for significant deviations of the cooling functions from a solar pattern are C, N, and O. (See Fig.~\ref{fig:abundance_spread} in Appendix for abundance spread plots)

$M_{\rm cool}$ and $P$ were calculated using the non standard cooling functions and  the  results of the simulations are presented in Fig.~\ref{fig:m_p_non_solar}.  The abundance pattern remains an important factor when considering SN feedback. We find that using non-standard cooling functions can result in changes in $P$ deposition by a factor of up to 25\%. Such a difference stems from the fact that non-solar abundances for a given metallicity usually provide less cooling, as we found out from our work with \jina~ (also can be seen in Fig.~\ref{fig:non_solar}, comparing the integrals of solar vs non-solar pattern curves). Less cooling efficiency provides for larger cooling time scales, meaning larger $M_{\rm cool}$ and momentum deposition. This is particularly important for the emergent field of simulations of galaxies at the epoch of reionization, where metallicities are still low and the abundance pattern is not necessarily solar.

\section{Discussion}
\label{sec:discussion}

In this Section, we discuss the implications of our results. We first discuss a few limitations of current models of SN feedback that do not fully take into account the evolution of SNRs in low metallicity environments. We propose a series of formulae that properly capture this physics and test them against cruder models using 3D simulations of galactic disc patches; in order to isolate the effect of metallicity, the abundance pattern is kept fixed in these tests. Finally, we extend the discussion to cases with non-solar abundance pattern and highlight how SNR measurements can be used to place constraints on the chemistry of the ISM.

\subsection{Solar ISM Abundance Fits}
\label{sec:fitting_formulae}

Although different abundance patterns yield significant differences in the cooling function, at low ISM metallicity $Z<10^{-2}\,Z_{\odot}$ the differences become smaller, because the contribution from metals to the cooling function becomes small compared to the contribution from H and He. As shown in Figure~\ref{fig:q_solar}, this implies that that the cooling mass and terminal momentum of a SNR evolving in a low metallicity ISM saturate to well defined values at metallicity $Z<10^{-2}\,Z_{\odot}$, and decrease at metallicity $Z>10^{-1}\,Z_{\odot}$. Since at low metallicity the evolution of SNRs is mostly determined by H and He cooling, the saturation in cooling mass and momentum depends weakly on the abundance pattern. However, the full dependence of SNR cooling mass and momentum on metallicity has not been taken into account in most models for SN feedback in galaxy formation simulations. As a matter of fact, several high-impact methods for momentum-driven SN feedback in cosmological zoom-in simulations use incomplete implementations of the physics of SNRs. For instance, \cite{Hopkins2014} take into account the dependence of SNR cooling radii on density and metallicity, but do not include a scaling of the terminal momentum with metallicity. On the other hand, \cite{Kimm2015} use a momentum floor for metallicity $Z<10^{-2}\,Z_{\odot}$ which only roughly reproduces the scaling of  Figure~\ref{fig:q_solar} in this paper. Power-law fitting formulae calibrated on 3-d simulations of SNRs were proposed by \cite{MartizziSubGrid2015}, with the caveat that the low metallicity saturation of cooling radius and terminal momentum are not included by construction and need to be enforced by hand.

In order to facilitate the implementation of physically-motivated sub-grid models of SNRs, we provide an improved model for $M_{\rm cool}$ and $P$ for future studies: a three parameter fit for Z/Z$_{\odot}$ which is valid  from $10^{-4}$ to $10^{0}$ using the curve fitting software {\sf ZunZun}\footnote{\url{www.zunzun.com}}. Its intent is to be easily incorporated into a sub-grid model in any existing cosmological code. The formulae naturally capture the low metallicity trend. The form is as follows:
\begin{eqnarray}
X = A\alpha^{\log_{10}\left(\frac{100}{n}\right)}\left(\frac{Z}{Z_{\odot}}-\beta\right)^{\gamma}
\end{eqnarray}
where $\alpha$, $\beta$, $\gamma$ are fitting parameters given in Table \ref{tb:fit_coef}, with A being a normalization factor based on the values at $Z/Z_{\odot}=10^0$, and n is the number density.

\begin{table}
\centering
\begin{tabular}{ccccc}
\hline
\hline
X & A & $\alpha$	&	$\beta$	&  $\gamma$	\\
\hline
$M_{\rm cool}$ & $1.177\times 10^{36}$  &   1.944	&	-0.057	 &	-0.302	\\
$P$   & $3.252\times 10^{43}$ &   1.404	&	-0.057	&	-0.149	\\
\hline
\hline
\end{tabular}
\caption{Fit coefficients for $M_{\rm cool}$ and $P$.}
\label{tb:fit_coef}
\end{table}

\begin{figure*}
\centering
\subfigure{\includegraphics[width=1\columnwidth]{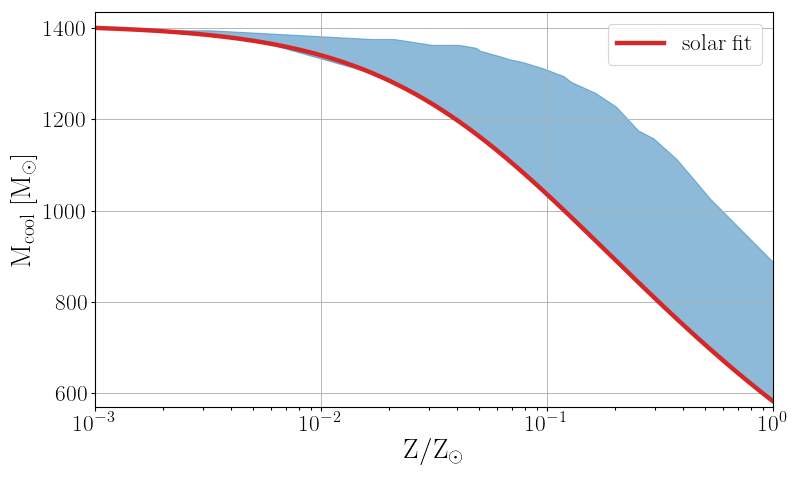}
\label{fig:H_zvM_5.3}}%
\subfigure{\includegraphics[width=1.05\columnwidth]{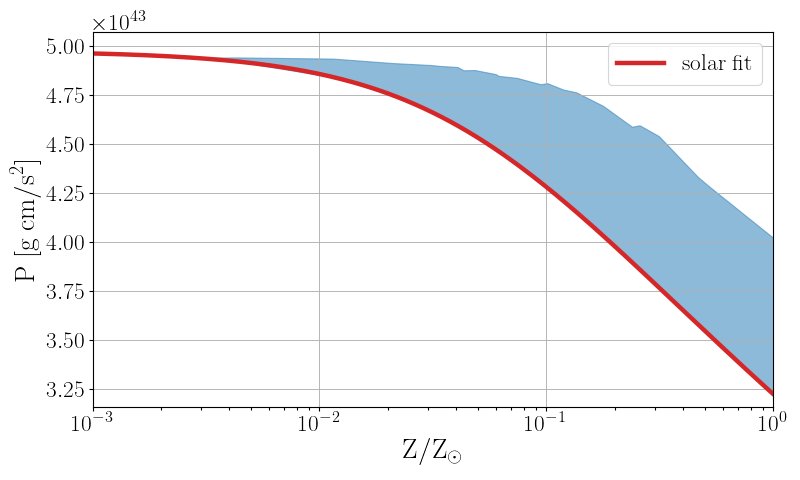}
\label{fig:H_zvM_6.5}}%
\caption{$M_{\rm cool}$ and $P$ as a function of relative metallicity plots, where the red curve is our fit for solar based ISM pattern results, and the shaded region contains the results from all the non-solar abundance ISM results, using \jina\ stars as tracer ISM particles. }
\label{fig:m_p_non_solar}
\end{figure*}

Fig.~\ref{fig:m_p_non_solar} presents the solar abundance based fit along with a shaded region containing the typical spread of $M_{\rm cool}$ and $P$ for the non-solar abundance results. This provides a clear upper limit to the validity of solar abundance cooling functions in effectively describing the cooling properties of the ISM.  We note that our results of metallicity going to 0 as M$_{\rm cool}$ approaches 1400 M$_{\odot}$ in Fig.~\ref{fig:m_p_non_solar} are consistent with analytical estimates \edit1{for} primordial gas composition by \cite{Efstathiou_2000}.

\subsection{Implication for Galaxy Evolution Simulations}
\label{sec:galev}

 In order to test the new fitting formulae that describe SNe feedback, we have run 3-d simulations of patches of vertically stratified ISM, performed with the \ramses\ adaptive mesh refinement code \citep{Teyssier2002}. The initial conditions and boundary conditions are the same used for the `ULTRA-MW' setup described in \citet{Martizzi2016}, and resemble those of a star-forming galactic disc with surface density $\Sigma_{\rm gas} \approx 50\, M_{\odot}/{\rm pc}^2$. In practice, gas in a
$1 \, {\rm kpc} \times 1 \, {\rm kpc} \times 1 \, {\rm kpc}$ box is initially set in hydrostatic equilibrium in an external gravitational potential that mimics the vertical stratification of a galactic disc:
\begin{equation}\label{eq:potential}
 \phi (z)=a_1\left[ \sqrt{z^2+z_{0}^2}-z_0\right]+\frac{a_2}{2}z^2,
\end{equation}
 where $z$ is the distance from the disc mid-plane, and $a_1 = 1.26 \times 10^{-2} \, {\rm kpc \, Myr^{-2}}$, $a_2 = 4.87 \times 10^{-3} \, {\rm Myr^{-2}}$, $z_0=180 \, {\rm pc}$. Gas self-gravity is not included, in order to isolate the effects of SN feedback. In fact, although being more realistic, simulations with self-gravity add significant non-linearity to the system, making the interpretation of the emergent processes caused by SN feedback more complicated \citep[e.g.]{2016MNRAS.456.3432G,2018ApJ...853..173K,Martizzi2020}. The simulations use a fixed Cartesian grid of size $256^3$ that covers a physical domain of size $1 \, {\rm kpc} \times 1 \, {\rm kpc} \times 1 \, {\rm kpc}$, i.e. the cell size is $\Delta x = 3.9 \, {\rm pc}$. Adaptive mesh refinement was not used for this setup, because one of the main motivations of the original paper was to to study turbulence in the simulated disc with homogeneous resolution throughout the computational box. Robustness tests of the numerical setup  are described in detail in \cite{Martizzi2016} and show excellent convergence as a function of numerical resolution; for this reason, these tests will not be repeated here and we refer the reader to the original paper.

The total metallicity of the gas is allowed to evolve over time, but the relative abundances of elements are fixed to the solar values. In this sense, as SNe explode and inject metal-rich ejecta the total metallicity can grow, but not the mixture of elements. The initial metallicity of the gas is set to $Z_0 = 10^{-6} \, Z_{\odot}$. Gas is not allowed to cool radiatively below a temperature $T_{\rm floor}=10^4 \, {\rm K}$, but it can reach lower temperatures by hydrodynamical processes, such as adiabatic expansion.

\begin{figure*}
\centering
\includegraphics[width=0.9\textwidth]{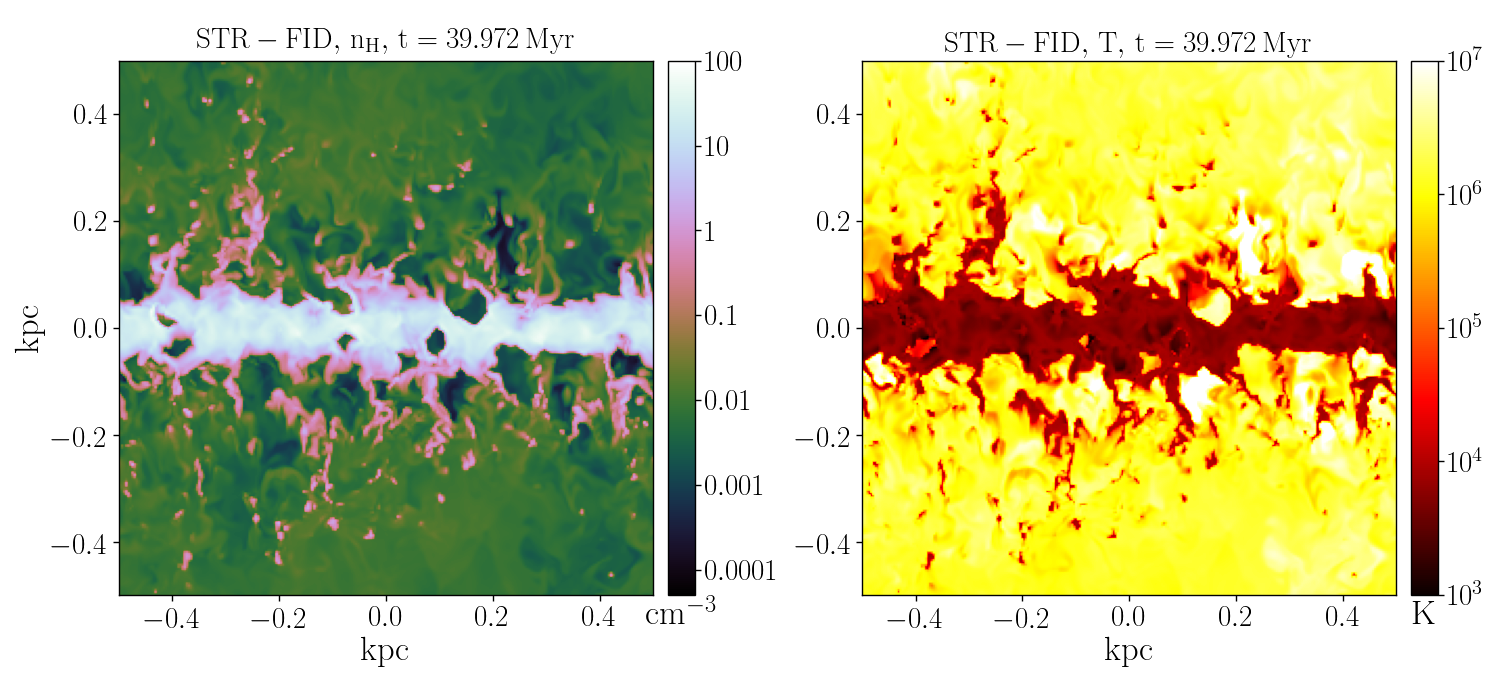}
\caption{Density (left) and temperature (right) maps of the final snapshot of the STR-FID simulation. The maps are extracted for a thin slice of thickness 8 pc passing through the box center. }
\label{fig:ramses}
\end{figure*}

\begin{table*}[t]
\centering
\makebox[\linewidth]{
\begin{tabular}{lcccc}
\hline
\hline
\multicolumn{5}{l}{Mass Loading of Galactic Outflows} \\
\hline
Simulation Name & $\eta_{\rm tot}(z=150 \, {\rm pc})$ & $\eta_{\rm tot}(z=495 \, {\rm pc})$ & $\eta_{\rm esc}(z=150 \, {\rm pc})$ & $\eta_{\rm esc}(z=495 \, {\rm pc})$ \\
\hline
STR-FID & 2.68 & $4.83\times 10^{-1}$ & $8.44\times 10^{-2}$ & $1.51\times 10^{-4}$ \\
STR-PL & 2.95 & $4.96\times 10^{-1}$ & $8.42\times 10^{-2}$ & $5.95\times 10^{-4}$ \\
STR-PL-FXZ & 5.73 & 1.16 & $1.91\times 10^{-1}$ & $1.92\times 10^{-2}$ \\
\hline
\hline
\end{tabular}
}
\caption{Mass loading factor of galactic outflows in simulations of stratified media performed with different models for SN injection. {\it Column 1:} simulation name. {\it Column 2:} $\eta_{\rm tot}$, the total mass loading of outflowing material, measured at height $z=150 \, {\rm pc}$ from the disc mid-plane. {\it Column 3:} $\eta_{\rm tot}$, the total mass loading of outflowing material, measured at height $z=495 \, {\rm pc}$ from the disc mid-plane. {\it Column 4:} $\eta_{\rm esc}$, the mass loading of material outflowing with velocity $|v_{\rm z}|\geq 300 \, {\rm km/s}$, measured at height $z=150 \, {\rm pc}$ from the disc mid-plane. {\it Column 5:} $\eta_{\rm esc}$, the mass loading of material outflowing with velocity $v\geq 300 \, {\rm km/s}$, measured at height $z=495 \, {\rm pc}$ from the disc mid-plane.} \label{tab:mass_loading}
\end{table*}

\begin{figure*}
\centering
\makebox[\textwidth][c]{\includegraphics[width=0.9\textwidth]{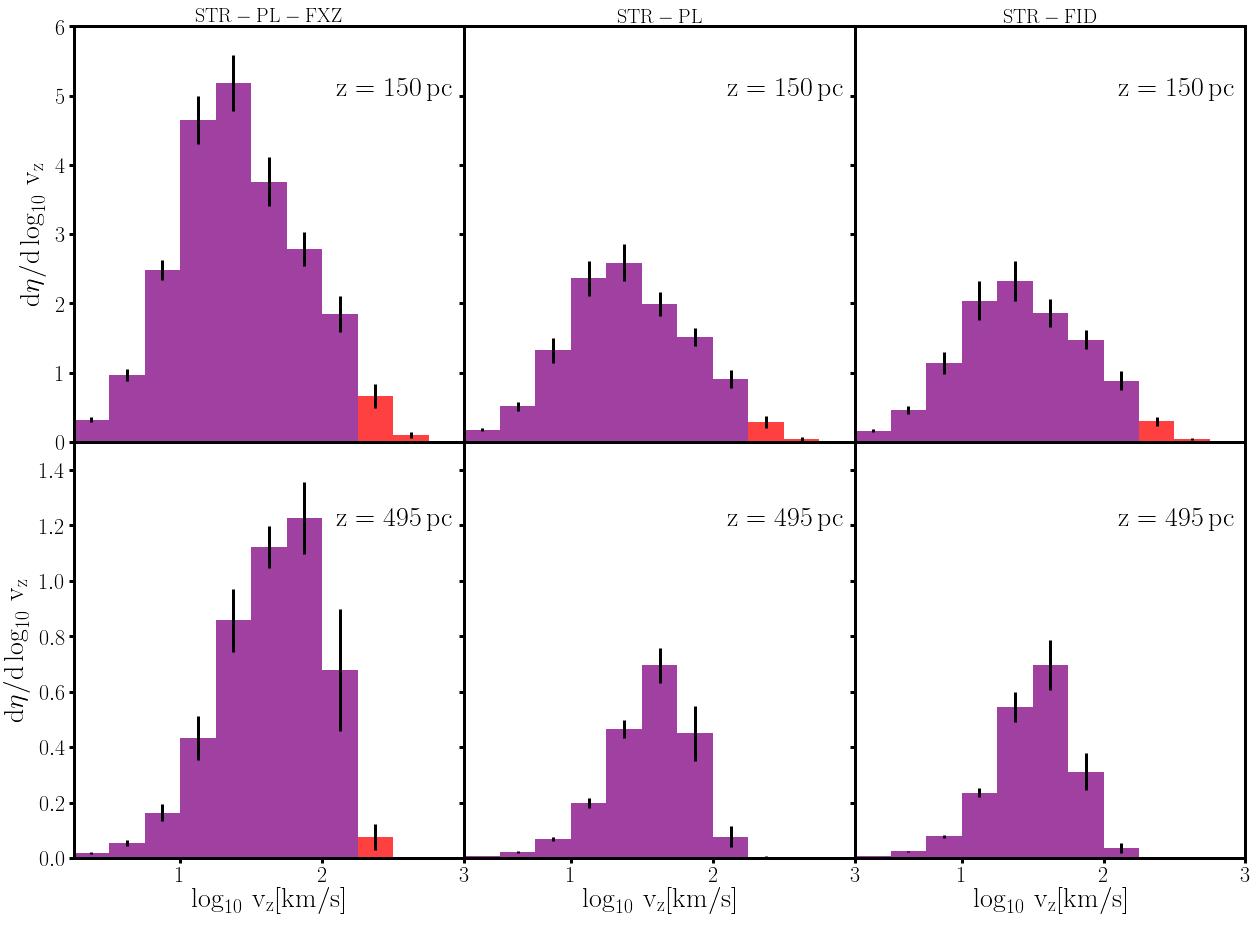}}
\caption{Histograms showing the mass loading of galactic outflows in simulations of SN feedback in stratified media as a function of velocity of the outflowing material. The integral of the histograms is the total mass loading at the height where the outflow is measured. Top: mass loading distributions measured in the region where the outflows are launched, at height $z=150 \, {\rm pc}$ from the disc mid-plane. Bottom: mass loading distributions measured at the edge of the computational box, at height $z=495 \, {\rm pc}$ from the disc mid-plane. The STR-PL-FXZ, which adopts power law fits for the SNR cooling mass and radial momentum and keeps the metallicity fixed at low values, overestimates the mass loading of galactic winds compared to the STR-FID and STR-PL simulations, which model the dependence of SNR evolution on ISM metallicity and density more accurately. Inclusion of the appropriate physical dependence of SNR cooling mass and radial momentum influences mass loading factors of galactic outflows by a factor $\approx 2$.}
\label{fig:mass_loading}
\end{figure*}

SNe are seeded at a fixed rate at random locations within a region extending 100 pc above and below the galactic disc mid-plane. The SN rate is chosen to be $3\times 10^{-4} \, {\rm SN/yr/kpc^3}$, which is set assuming that there is $\approx 1$ SN per $100 \, M_{\odot}$ of newly formed stars and that the star formation rate surface density is given by the upper envelope of the star formation law of \cite{Kennicutt2007}. Each time a SN goes off, it deposits its ejecta mass ($M_{\rm ej}=3 \, M_{\odot}$) within a sphere of radius $R_{\rm ej}= 8 \, {\rm pc}$. A fraction $0.4$ of the ejecta mass is assumed to be made of metals, which are uniformly distributed over the injection sphere. To compare with other feedback models, radial momentum and thermal energy are also deposited in the same sphere using an adaptive sub-resolution model inspired by \cite{MartizziSubGrid2015} and updated in this work. This method is based on knowledge of the functional dependence of the scaling of SNR cooling mass and radial momentum with density and metallicity. To highlight the effect of our new fitting formulae for cooling SNRs (see Section \ref{sec:fitting_formulae}), we have performed three test runs:
\begin{enumerate}
\item The STR-FID simulation, our fiducial run, adopts our new formulae for SN injection (Section \ref{sec:fitting_formulae}) and includes the full dependence of cooling mass and momentum on metallicity and density. The new formulae automatically include the lack of dependence of the cooling mass and momentum at metallicity $Z< 0.01 \, Z_{\odot}$, due to the fact that at such low metallicity the only relevant coolants are H and He.
\item The STR-PL simulation uses the the power-law formulae of \cite{MartizziSubGrid2015} for $Z\geq 0.01 \, Z_{\odot}$, but uses the values for the cooling radius at metallicity $Z=0.01\,Z_{\odot}$ when $Z<0.01 \, Z_{\odot}$. This approach allows us to include the fact that at low metallicity, the only coolants are H and He.
\item The STR-PL-FXZ simulation uses the the power-law formulae of \cite{MartizziSubGrid2015} extrapolated down to metallicity $Z=Z_{0}=10^{-6} \, Z_{\odot}$. The metallicity is kept fixed at $Z=Z_{0}=10^{-6} \, Z_{\odot}$ at all times. In practice, this is the wrong way of implementing the formulae of \cite{MartizziSubGrid2015}, which should not be extrapolated to arbitrary low metallicities.
\end{enumerate}

The numerical setups described above are not intended to capture ultra-realistic configurations in specific galaxies, but they are designed to isolate the effects of SN feedback in a controlled, idealised environment. In order to model the whole range of physical processes involved in the evolution of realistic galaxies, self-gravity, star formation, SN time delays, cooling functions from molecules at $T<10^4 \, {\rm K}$, heating processes, full disc geometry and the presence of a hot gaseous halo. Including all these physical processes, is beyond the scope of this discussion, and does not allow to isolate the physics of the evolution of SNRs in the ISM. With our setup, it is possible to highlight differences between different SN seeding models and isolate the dependence of SN feedback on metallicity in the non-linear multiple-SN regime, which is not probed by simulations of isolated SNRs. Additional advantages of the setup are that (I) radiative cooling is included in the appropriate range of temperatures at which SNRs cooling in the ISM occurs, and (II) the temperature floor of $10^4 \, {\rm K}$ mimics the balance of cooling and heating sources in the dense ISM.

Fig.~\ref{fig:ramses} shows density and temperature maps extracted from thin slices passing through the computational box center and at the final time of the STR-FID simulation. The vertical stratification, as well as the turbulent nature of the simulated ISM is evident from the slices. We omit density and temperature maps from the STR-PL and STR-PL-FXZ simulations, because they look qualitatively similar to the STR-FID run. These simulations develop into a steady galactic wind with a total mass loading $\eta = \dot{M}_{\rm out}/{\rm SFR}\approx 1$.

In their discussion, \cite{Martizzi2016} show that outflows generated by SN feedback in locally stratified simulations do not fully develop into supersonic outflows characteristic of galactic winds. However, the energetics and mass loading of these winds {\itshape at their launching height} ($\approx 150 \, {\rm pc}$, i.e. the thermal scale height; above which thermal energy is rapidly converted into wind kinetic energy) is accurately captured by the simulations. For this reason, examining the properties of such winds in the simulations presented in this paper constitutes a relevant test of the SN seeding models.

We highlight the differences made by the choice for the fitting formulae for SNR cooling mass and momentum on the mass loading of galactic outflows in Fig.~\ref{fig:mass_loading}. Each panel of this figure shows a histogram of the mass loading of material moving at a given velocity. Only outflowing gas is included in the calculation of the mass loading factor. The histograms are normalized to the total mass loading factor of the gas in each simulation and at the height where the outflow is measured. Results are shown for the wind at heights $z=150 \, {\rm pc}$ (the launching point of the wind, top panels) and $z=495 \, {\rm pc}$ (shown for completeness, bottom panels) from the disc mid-plane. The mass loading factors of the STR-PL-FXZ run are approximately twice as large as in the other two runs. The main reason is that extrapolating the \citet{MartizziSubGrid2015} power-law dependence of cooling mass and momentum on metallicity down to $Z=Z_{0}=10^{-6} \, Z_{\odot}$, produces an overestimate of the effects of a SN feedback (cooling is assumed to be less efficient than it actually is). STR-PL which uses the power-laws of \citet{MartizziSubGrid2015} without extrapolating them to low metallicity and STR-FID appear to be in very good agreement with each other, with minor differences. The new formulae proposed in this paper have an advantage over power-law fits, because they smoothly capture the physics of the problem even at metallicity $Z<0.01\, Z_{\odot}$ (see Fig.~\ref{fig:Mcool_solar} and Fig.~\ref{fig:Momentum_solar}).

Table~\ref{tab:mass_loading} summarizes the mass loading factors at heights $z=150 \, {\rm pc}$ and $z=495 \, {\rm pc}$ from the disc mid-plane. We include both the total mass loading factor of all the outflowing material at each height, $\eta_{\rm tot}$, and the mass loading factor of outflowing gas with velocity $|v_{\rm z}|\geq 300 \, {\rm km/s}$ at each height, $\eta_{\rm esc}$. The latter represents the mass loading of the material that has velocity higher than the typical escape velocity of a massive galaxy, i.e. the gas that will ultimately be ejected without being re-accreted onto the galaxy. The table offers quantitative confirmation that the STR-PL-FXZ simulation overestimates the outflow mass loading by a factor $\approx 2$ with respect to STR-PL and STR-FID.

\subsection{Effects of the non-solar ISM abundance pattern}

\begin{figure*}
\centering
\subfigure{\includegraphics[width=1\columnwidth]{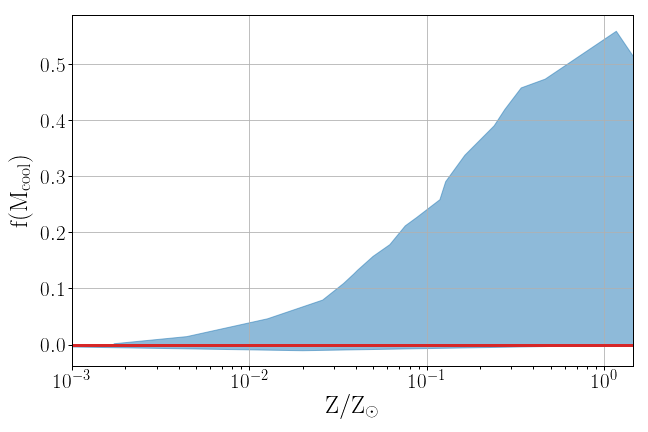}}%
\subfigure{\includegraphics[width=1\columnwidth]{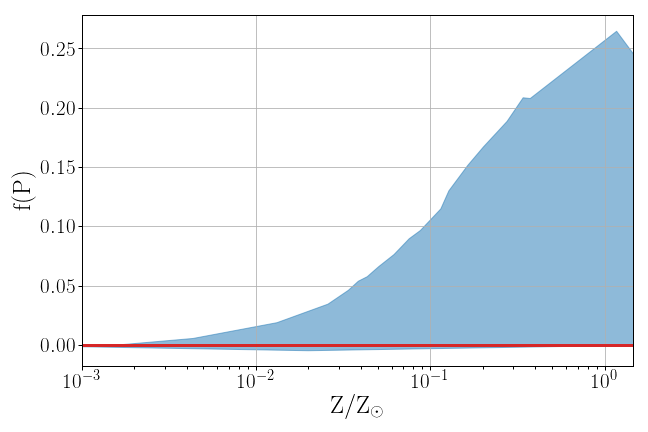}}%
\caption{Fractional differences of the quantities due to the variation in the abundance-pattern at a given metallicity. Y-axis is defined as $f(x)=\frac{x-x_{\sun}}{x_{\sun}}$. We found the maximum deviations in $\rm M_{cool}$ to be $\approx 55\%$ and in P $\approx 27\%$.}
\label{fig:delta_plots}
\end{figure*}

Comparing the solar-abundance pattern results to the non-solar has yielded significant deviations of up to $\approx 55\%$ in $M_{\rm cool}$ and up to $\approx 27\%$ in $P$. These findings are summarized in Fig.~\ref{fig:delta_plots}. It is important to note that while the maximum deviations from the solar pattern  arise in the early Universe when the gas is not well mixed, they don't play a role in the cooling of the ISM until after the overall metallicity reaches values of $Z>10^{-2}\,Z_\sun$.

It is also important to note that these results represent an upper bound for maximum effect of the swept-up ISM abundance onto the evolution of the SNR. The primary assumption that we took leading to this is using the stellar atmosphere composition measurements from \jina\ as tracer particles of the ISM composition from which they have been formed. In reality, the primordial ISM should be less metal abundant, as the stars measured have been evolving and potentially mass transferring, hence enriching their surfaces beyond  their birth metallicity. The approach chosen here follows our intention to quantify the maximum potential uncertainty of using solar abundance pattern to study the stellar formation and galaxy evolution through simulations.

\section{Conclusion}
\label{sec:conclusion}
For the study of low-$Z$ subgrid SN feedback models, assuming solar abundance pattern, we first developed accurate fits for $M_{\rm cool}$ and feedback, based on the suite of \flash\ runs performed. This was done to test the typical assumption of a power-law fit for stellar feedback, that overestimates feedback at low-$Z$. The fits used can be found in Section \ref{sec:fitting_formulae} and be used as a subgrid SN feedback model within any galaxy evolution simulation. Next, we performed a comparison study of power-law-based and the new feedback subgrid models within galaxy evolution simulations using \ramses. These simulations  yielded a decrease in mass loading factor (mass outflow) within the galactic disk by a factor of $\approx 2$ (Fig.~\ref{fig:mass_loading}) when the new fits were used, hence eliminating the overestimation of SN feedback at low-Z. In principle, less outflow, i.e. less feedback, may directly lead to higher star formation rates in the early Universe. Reduced feedback and higher star formation rates in primordial galaxies may help alleviate the {\it Impossibly Early Galaxies} problem \citep{2016ApJ...824...21S} in which too many massive quenched galaxies appear to be found at redshifts $z=4-8$ \citep{2004Natur.430..181G,Duncan2014,Bouwens2016} compared to the expectations of $\Lambda$ Cold Dark Matter cosmology: in fact, primordial galaxies with weaker feedback and higher star formation rates may run out of gas for star formation and quench earlier than expected. This hypothesis needs to be tested with dedicated cosmological simulations.

Considering the widely used assumption of solar abundance pattern in cosmological simulations, while metallicity is enriched as a function of time, we've tested the effects of non-solar abundance pattern on SN feedback. Since even in the early Universe, ejecta metals do not affect SNR cooling, with the swept-up ISM material dominating the energy losses (Fig.~\ref{fig:Fe_pristine}), we focused on testing various ISM abundance regimes to get the maximum deviations in SN feedback by accounting for non-solar abundance patterns. Metal-poor stars from \jina\ were used as tracer particles, hence setting the maximum variations expected in ISM abundances. M$_{\rm cool}$ was found to differ by $\approx 55\%$ and feedback by $\approx 27\%$, growing with  increasing metallicity (Fig.~\ref{fig:delta_plots}).

Our study of SNRs in an ISM with non-solar abundance pattern constitutes a significant improvement for models of SN feedback in metal-poor primordial galaxies, but several aspects of this problem still need to be investigated in detail. For instance, SNR dynamics in an inhomogeneous and turbulent medium can be significantly different than in a homogeneous medium \citep[e.g.][]{MartizziSubGrid2015,2015ApJ...802...99K,2019MNRAS.482.1602Z}. Joint studies of the degree of inhomogeneity in the primordial ISM, non-solar abundance patterns, and SN feedback are currently missing. Furthermore, if the SN rate is sufficiently high in a certain region of the ISM, clustering of SNe in time and space can lead to the formation of hot superbubbles and enhanced feedback  \citep[e.g.][]{2017ApJ...834...25K,2018MNRAS.481.3325F,2019MNRAS.483.3647G}, phenomena that are currently not explicitly taken into account in our study. These issues need to be effectively  tackled in future work.\\

\section{Acknowledgment}
We would like to thank the referee for  useful comments and Dongwook Lee for help and support with setting up our models in \flash.  The software used in this work was in part developed by the DOE NNSA-ASC OASCR Flash Center at the University of Chicago. UCSC grant. DM was supported by the CTA and DARK-Carlsberg Foundation Fellowship. DM acknowledges contribution from the Danish council for independent research under the project ``Fundamentals of Dark Matter Structures'', DFF - 6108-00470. AK, DM, PM and ERR acknowledge support by the Danish National Research Foundation (DNRF132). Part of the simulations used in this paper were performed on the University of Copenhagen high-performance computing cluster funded by a grant from VILLUM FONDEN (project number 16599).

\software{FLASH (\cite{Fryxell2000}), RAMSES (\cite{Teyssier2002}), CoolingCurve (\url{https://github.com/pikarpov/CoolingCurve}), ZunZun (\url{www.zunzun.com})}

\bibliographystyle{apj}
\bibliography{bibliography}

\begin{thebibliography}{}
\expandafter\ifx\csname natexlab\endcsname\relax\def\natexlab#1{#1}\fi
\providecommand{\url}[1]{\href{#1}{#1}}
\providecommand{\dodoi}[1]{doi:~\href{http://doi.org/#1}{\nolinkurl{#1}}}
\providecommand{\doeprint}[1]{\href{http://ascl.net/#1}{\nolinkurl{http://ascl.net/#1}}}
\providecommand{\doarXiv}[1]{\href{https://arxiv.org/abs/#1}{\nolinkurl{https://arxiv.org/abs/#1}}}

\bibitem[{{Abohalima} \& {Frebel}(2018)}]{JINAbase2018}
{Abohalima}, A., \& {Frebel}, A. 2018, \apjs, 238, 36,
  \dodoi{10.3847/1538-4365/aadfe9}

\bibitem[{{Bouwens} {et~al.}(2016){Bouwens}, {Oesch}, {Labb{\'e}},
  {Illingworth}, {Fazio}, {Coe}, {Holwerda}, {Smit}, {Stefanon}, {van Dokkum},
  {Trenti}, {Ashby}, {Huang}, {Spitler}, {Straatman}, {Bradley}, \&
  {Magee}}]{Bouwens2016}
{Bouwens}, R.~J., {Oesch}, P.~A., {Labb{\'e}}, I., {et~al.} 2016, \apj, 830,
  67, \dodoi{10.3847/0004-637X/830/2/67}

\bibitem[{{Bryans} {et~al.}(2009){Bryans}, {Landi}, \& {Savin}}]{Bryans2009}
{Bryans}, P., {Landi}, E., \& {Savin}, D.~W. 2009, \apj, 691, 1540,
  \dodoi{10.1088/0004-637X/691/2/1540}

\bibitem[{{Cioffi} {et~al.}(1988){Cioffi}, {McKee}, \&
  {Bertschinger}}]{Cioffi1988}
{Cioffi}, D.~F., {McKee}, C.~F., \& {Bertschinger}, E. 1988, \apj, 334, 252,
  \dodoi{10.1086/166834}

\bibitem[{{Cox}(1972)}]{cox1972}
{Cox}, D.~P. 1972, \apj, 178, 159, \dodoi{10.1086/151775}

\bibitem[{{Dav{\'e}} {et~al.}(2016){Dav{\'e}}, {Thompson}, \&
  {Hopkins}}]{MUFASA2016}
{Dav{\'e}}, R., {Thompson}, R., \& {Hopkins}, P.~F. 2016, \mnras, 462, 3265,
  \dodoi{10.1093/mnras/stw1862}

\bibitem[{{Dorfi} \& {Voelk}(1996)}]{Dorfi1996}
{Dorfi}, E.~A., \& {Voelk}, H.~J. 1996, \aap, 307, 715

\bibitem[{{Dubois} {et~al.}(2014){Dubois}, {Pichon}, {Welker}, {Le Borgne},
  {Devriendt}, {Laigle}, {Codis}, {Pogosyan}, {Arnouts}, {Benabed}, {Bertin},
  {Blaizot}, {Bouchet}, {Cardoso}, {Colombi}, {de Lapparent}, {Desjacques},
  {Gavazzi}, {Kassin}, {Kimm}, {McCracken}, {Milliard}, {Peirani}, {Prunet},
  {Rouberol}, {Silk}, {Slyz}, {Sousbie}, {Teyssier}, {Tresse}, {Treyer},
  {Vibert}, \& {Volonteri}}]{Dubois2014}
{Dubois}, Y., {Pichon}, C., {Welker}, C., {et~al.} 2014, \mnras, 444, 1453,
  \dodoi{10.1093/mnras/stu1227}

\bibitem[{{Duncan} {et~al.}(2014){Duncan}, {Conselice}, {Mortlock}, {Hartley},
  {Guo}, {Ferguson}, {Dav{\'e}}, {Lu}, {Ownsworth}, {Ashby}, {Dekel},
  {Dickinson}, {Faber}, {Giavalisco}, {Grogin}, {Kocevski}, {Koekemoer},
  {Somerville}, \& {White}}]{Duncan2014}
{Duncan}, K., {Conselice}, C.~J., {Mortlock}, A., {et~al.} 2014, \mnras, 444,
  2960, \dodoi{10.1093/mnras/stu1622}

\bibitem[{Efstathiou(2000)}]{Efstathiou_2000}
Efstathiou, G. 2000, Monthly Notices of the Royal Astronomical Society, 317,
  697–719, \dodoi{10.1046/j.1365-8711.2000.03665.x}

\bibitem[{{Faucher-Gigu{\`e}re} {et~al.}(2013){Faucher-Gigu{\`e}re},
  {Quataert}, \& {Hopkins}}]{Faucher2013}
{Faucher-Gigu{\`e}re}, C.-A., {Quataert}, E., \& {Hopkins}, P.~F. 2013, \mnras,
  433, 1970, \dodoi{10.1093/mnras/stt866}

\bibitem[{{Fielding} {et~al.}(2018){Fielding}, {Quataert}, \&
  {Martizzi}}]{2018MNRAS.481.3325F}
{Fielding}, D., {Quataert}, E., \& {Martizzi}, D. 2018, \mnras, 481, 3325,
  \dodoi{10.1093/mnras/sty2466}

\bibitem[{{Frebel} \& {Norris}(2013)}]{Frebel2013}
{Frebel}, A., \& {Norris}, J.~E. 2013, {Metal-Poor Stars and the Chemical
  Enrichment of the Universe}, ed. T.~D. {Oswalt} \& G.~{Gilmore}, Vol.~5, 55

\bibitem[{{Fryxell} {et~al.}(2000){Fryxell}, {Olson}, {Ricker}, {Timmes},
  {Zingale}, {Lamb}, {MacNeice}, {Rosner}, {Truran}, \& {Tufo}}]{Fryxell2000}
{Fryxell}, B., {Olson}, K., {Ricker}, P., {et~al.} 2000, \apjs, 131, 273,
  \dodoi{10.1086/317361}

\bibitem[{{Gentry} {et~al.}(2019){Gentry}, {Krumholz}, {Madau}, \&
  {Lupi}}]{2019MNRAS.483.3647G}
{Gentry}, E.~S., {Krumholz}, M.~R., {Madau}, P., \& {Lupi}, A. 2019, \mnras,
  483, 3647, \dodoi{10.1093/mnras/sty3319}

\bibitem[{{Girichidis} {et~al.}(2016){Girichidis}, {Walch}, {Naab}, {Gatto},
  {W{\"u}nsch}, {Glover}, {Klessen}, {Clark}, {Peters}, {Derigs}, \&
  {Baczynski}}]{2016MNRAS.456.3432G}
{Girichidis}, P., {Walch}, S., {Naab}, T., {et~al.} 2016, \mnras, 456, 3432,
  \dodoi{10.1093/mnras/stv2742}

\bibitem[{{Glazebrook} {et~al.}(2004){Glazebrook}, {Abraham}, {McCarthy},
  {Savaglio}, {Chen}, {Crampton}, {Murowinski}, {J{\o}rgensen}, {Roth}, {Hook},
  {Marzke}, \& {Carlberg}}]{2004Natur.430..181G}
{Glazebrook}, K., {Abraham}, R.~G., {McCarthy}, P.~J., {et~al.} 2004, \nat,
  430, 181, \dodoi{10.1038/nature02667}

\bibitem[{Gnat \& Ferland(2012)}]{Gnat2012}
Gnat, O., \& Ferland, G.~J. 2012, The Astrophysical Journal Supplement Series,
  199, 20, \dodoi{10.1088/0067-0049/199/1/20}

\bibitem[{{Herwig}(2005)}]{Herwig2005}
{Herwig}, F. 2005, \araa, 43, 435,
  \dodoi{10.1146/annurev.astro.43.072103.150600}

\bibitem[{{Hopkins} {et~al.}(2014){Hopkins}, {Kere{\v s}}, {O{\~n}orbe},
  {Faucher-Gigu{\`e}re}, {Quataert}, {Murray}, \& {Bullock}}]{Hopkins2014}
{Hopkins}, P.~F., {Kere{\v s}}, D., {O{\~n}orbe}, J., {et~al.} 2014, \mnras,
  445, 581, \dodoi{10.1093/mnras/stu1738}

\bibitem[{{Joung} \& {Mac Low}(2006)}]{Joung2006}
{Joung}, M.~K.~R., \& {Mac Low}, M.-M. 2006, \apj, 653, 1266,
  \dodoi{10.1086/508795}

\bibitem[{{Kennicutt} {et~al.}(2007){Kennicutt}, {Calzetti}, {Walter}, {Helou},
  {Hollenbach}, {Armus}, {Bendo}, {Dale}, {Draine}, {Engelbracht}, {Gordon},
  {Prescott}, {Regan}, {Thornley}, {Bot}, {Brinks}, {de Blok}, {de Mello},
  {Meyer}, {Moustakas}, {Murphy}, {Sheth}, \& {Smith}}]{Kennicutt2007}
{Kennicutt}, Jr., R.~C., {Calzetti}, D., {Walter}, F., {et~al.} 2007, \apj,
  671, 333, \dodoi{10.1086/522300}

\bibitem[{{Kim} \& {Ostriker}(2015)}]{2015ApJ...802...99K}
{Kim}, C.-G., \& {Ostriker}, E.~C. 2015, \apj, 802, 99,
  \dodoi{10.1088/0004-637X/802/2/99}

\bibitem[{{Kim} \& {Ostriker}(2018)}]{2018ApJ...853..173K}
---. 2018, \apj, 853, 173, \dodoi{10.3847/1538-4357/aaa5ff}

\bibitem[{{Kim} {et~al.}(2017){Kim}, {Ostriker}, \&
  {Raileanu}}]{2017ApJ...834...25K}
{Kim}, C.-G., {Ostriker}, E.~C., \& {Raileanu}, R. 2017, \apj, 834, 25,
  \dodoi{10.3847/1538-4357/834/1/25}

\bibitem[{{Kimm} \& {Cen}(2014)}]{Kimm2014}
{Kimm}, T., \& {Cen}, R. 2014, \apj, 788, 121,
  \dodoi{10.1088/0004-637X/788/2/121}

\bibitem[{{Kimm} {et~al.}(2015){Kimm}, {Cen}, {Devriendt}, {Dubois}, \&
  {Slyz}}]{Kimm2015}
{Kimm}, T., {Cen}, R., {Devriendt}, J., {Dubois}, Y., \& {Slyz}, A. 2015,
  \mnras, 451, 2900, \dodoi{10.1093/mnras/stv1211}

\bibitem[{{Kobayashi} {et~al.}(2006){Kobayashi}, {Umeda}, {Nomoto}, {Tominaga},
  \& {Ohkubo}}]{Kobayashi2006}
{Kobayashi}, C., {Umeda}, H., {Nomoto}, K., {Tominaga}, N., \& {Ohkubo}, T.
  2006, \apj, 653, 1145, \dodoi{10.1086/508914}

\bibitem[{Komiya(2011)}]{Komiya2011}
Komiya, Y. 2011, The Astrophysical Journal, 736, 73,
  \dodoi{10.1088/0004-637x/736/1/73}

\bibitem[{{Krumholz} \& {McKee}(2005)}]{Krumholz2005}
{Krumholz}, M.~R., \& {McKee}, C.~F. 2005, \apj, 630, 250,
  \dodoi{10.1086/431734}

\bibitem[{{Lopez} \& {Fesen}(2018)}]{Lopez2018}
{Lopez}, L.~A., \& {Fesen}, R.~A. 2018, \ssr, 214, 44,
  \dodoi{10.1007/s11214-018-0481-x}

\bibitem[{{Lopez} {et~al.}(2011){Lopez}, {Ramirez-Ruiz}, {Huppenkothen},
  {Badenes}, \& {Pooley}}]{Lopez2011}
{Lopez}, L.~A., {Ramirez-Ruiz}, E., {Huppenkothen}, D., {Badenes}, C., \&
  {Pooley}, D.~A. 2011, \apj, 732, 114, \dodoi{10.1088/0004-637X/732/2/114}

\bibitem[{{Martizzi}(2019)}]{2019arXiv190710623M}
{Martizzi}, D. 2019, arXiv e-prints, arXiv:1907.10623.
\newblock \doarXiv{1907.10623}

\bibitem[{{Martizzi} {et~al.}(2015){Martizzi}, {Faucher-Gigu{\`e}re}, \&
  {Quataert}}]{MartizziSubGrid2015}
{Martizzi}, D., {Faucher-Gigu{\`e}re}, C.-A., \& {Quataert}, E. 2015, \mnras,
  450, 504, \dodoi{10.1093/mnras/stv562}

\bibitem[{{Martizzi} {et~al.}(2016){Martizzi}, {Fielding},
  {Faucher-Gigu{\`e}re}, \& {Quataert}}]{Martizzi2016}
{Martizzi}, D., {Fielding}, D., {Faucher-Gigu{\`e}re}, C.-A., \& {Quataert}, E.
  2016, \mnras, 459, 2311, \dodoi{10.1093/mnras/stw745}

\bibitem[{{McCray} \& {Fransson}(2016)}]{McCray2016}
{McCray}, R., \& {Fransson}, C. 2016, \araa, 54, 19,
  \dodoi{10.1146/annurev-astro-082615-105405}

\bibitem[{{McKee} \& {Ostriker}(1977)}]{McKee1977}
{McKee}, C.~F., \& {Ostriker}, J.~P. 1977, \apj, 218, 148,
  \dodoi{10.1086/155667}

\bibitem[{{Naiman} {et~al.}(2018){Naiman}, {Pillepich}, {Springel},
  {Ramirez-Ruiz}, {Torrey}, {Vogelsberger}, {Pakmor}, {Nelson}, {Marinacci},
  {Hernquist}, {Weinberger}, \& {Genel}}]{2018MNRAS.477.1206N}
{Naiman}, J.~P., {Pillepich}, A., {Springel}, V., {et~al.} 2018, \mnras, 477,
  1206, \dodoi{10.1093/mnras/sty618}

\bibitem[{Nomoto {et~al.}(2013)Nomoto, Kobayashi, \&
  Tominaga}]{annurev-astro-082812-140956}
Nomoto, K., Kobayashi, C., \& Tominaga, N. 2013, Annual Review of Astronomy and
  Astrophysics, 51, 457, \dodoi{10.1146/annurev-astro-082812-140956}

\bibitem[{Placco {et~al.}(2014)Placco, Frebel, Beers, \&
  Stancliffe}]{Placco2014}
Placco, V.~M., Frebel, A., Beers, T.~C., \& Stancliffe, R.~J. 2014, The
  Astrophysical Journal, 797, 21, \dodoi{10.1088/0004-637x/797/1/21}

\bibitem[{{Semenov} {et~al.}(2017){Semenov}, {Kravtsov}, \&
  {Gnedin}}]{Semenov2017}
{Semenov}, V.~A., {Kravtsov}, A.~V., \& {Gnedin}, N.~Y. 2017, \apj, 845, 133,
  \dodoi{10.3847/1538-4357/aa8096}

\bibitem[{{Shen} {et~al.}(2015){Shen}, {Cooke}, {Ramirez-Ruiz}, {Madau},
  {Mayer}, \& {Guedes}}]{2015ApJ...807..115S}
{Shen}, S., {Cooke}, R.~J., {Ramirez-Ruiz}, E., {et~al.} 2015, \apj, 807, 115,
  \dodoi{10.1088/0004-637X/807/2/115}

\bibitem[{{Shigeyama} \& {Tsujimoto}(1998)}]{Shigeyama1998}
{Shigeyama}, T., \& {Tsujimoto}, T. 1998, \apj, 507, L135,
  \dodoi{10.1086/311699}

\bibitem[{Simpson {et~al.}(2015)Simpson, Bryan, Hummels, \&
  Ostriker}]{Simpson2015}
Simpson, C.~M., Bryan, G.~L., Hummels, C., \& Ostriker, J.~P. 2015, The
  Astrophysical Journal, 809, 69, \dodoi{10.1088/0004-637x/809/1/69}

\bibitem[{Sneden {et~al.}(2008)Sneden, Cowan, \& Gallino}]{Sneden2008}
Sneden, C., Cowan, J.~J., \& Gallino, R. 2008, Annual Review of Astronomy and
  Astrophysics, 46, 241, \dodoi{10.1146/annurev.astro.46.060407.145207}

\bibitem[{{Steinhardt} {et~al.}(2016){Steinhardt}, {Capak}, {Masters}, \&
  {Speagle}}]{2016ApJ...824...21S}
{Steinhardt}, C.~L., {Capak}, P., {Masters}, D., \& {Speagle}, J.~S. 2016,
  \apj, 824, 21, \dodoi{10.3847/0004-637X/824/1/21}

\bibitem[{{Sutherland} \& {Dopita}(1993)}]{SutherlandDopita1993}
{Sutherland}, R.~S., \& {Dopita}, M.~A. 1993, \apjs, 88, 253,
  \dodoi{10.1086/191823}

\bibitem[{{Tang} \& {Wang}(2005)}]{Tang2005}
{Tang}, S., \& {Wang}, Q.~D. 2005, \apj, 628, 205, \dodoi{10.1086/430875}

\bibitem[{{Teyssier}(2002)}]{Teyssier2002}
{Teyssier}, R. 2002, \aap, 385, 337, \dodoi{10.1051/0004-6361:20011817}

\bibitem[{Thornton {et~al.}(1998)Thornton, Gaudlitz, Janka, \&
  Steinmetz}]{Thornton1998}
Thornton, K., Gaudlitz, M., Janka, H.-T., \& Steinmetz, M. 1998, The
  Astrophysical Journal, 500, 95, \dodoi{10.1086/305704}

\bibitem[{{Veilleux} {et~al.}(2005){Veilleux}, {Cecil}, \&
  {Bland-Hawthorn}}]{Veilleux2005}
{Veilleux}, S., {Cecil}, G., \& {Bland-Hawthorn}, J. 2005, \araa, 43, 769,
  \dodoi{10.1146/annurev.astro.43.072103.150610}

\bibitem[{{Vogelsberger} {et~al.}(2013){Vogelsberger}, {Genel}, {Sijacki},
  {Torrey}, {Springel}, \& {Hernquist}}]{Vogelsberger2013}
{Vogelsberger}, M., {Genel}, S., {Sijacki}, D., {et~al.} 2013, \mnras, 436,
  3031, \dodoi{10.1093/mnras/stt1789}

\bibitem[{{Vogelsberger} {et~al.}(2014){Vogelsberger}, {Genel}, {Springel},
  {Torrey}, {Sijacki}, {Xu}, {Snyder}, {Nelson}, \&
  {Hernquist}}]{Vogelsberger2014}
{Vogelsberger}, M., {Genel}, S., {Springel}, V., {et~al.} 2014, \mnras, 444,
  1518, \dodoi{10.1093/mnras/stu1536}

\bibitem[{{Zhang} \& {Chevalier}(2019)}]{2019MNRAS.482.1602Z}
{Zhang}, D., \& {Chevalier}, R.~A. 2019, \mnras, 482, 1602,
  \dodoi{10.1093/mnras/sty2769}

\end{thebibliography}

\appendix
\section{Resolution Study}
\label{sec:res}
\begin{figure*}[!h]
\centering
\subfigure[Thermal Energy]{\includegraphics[width=0.5\columnwidth]{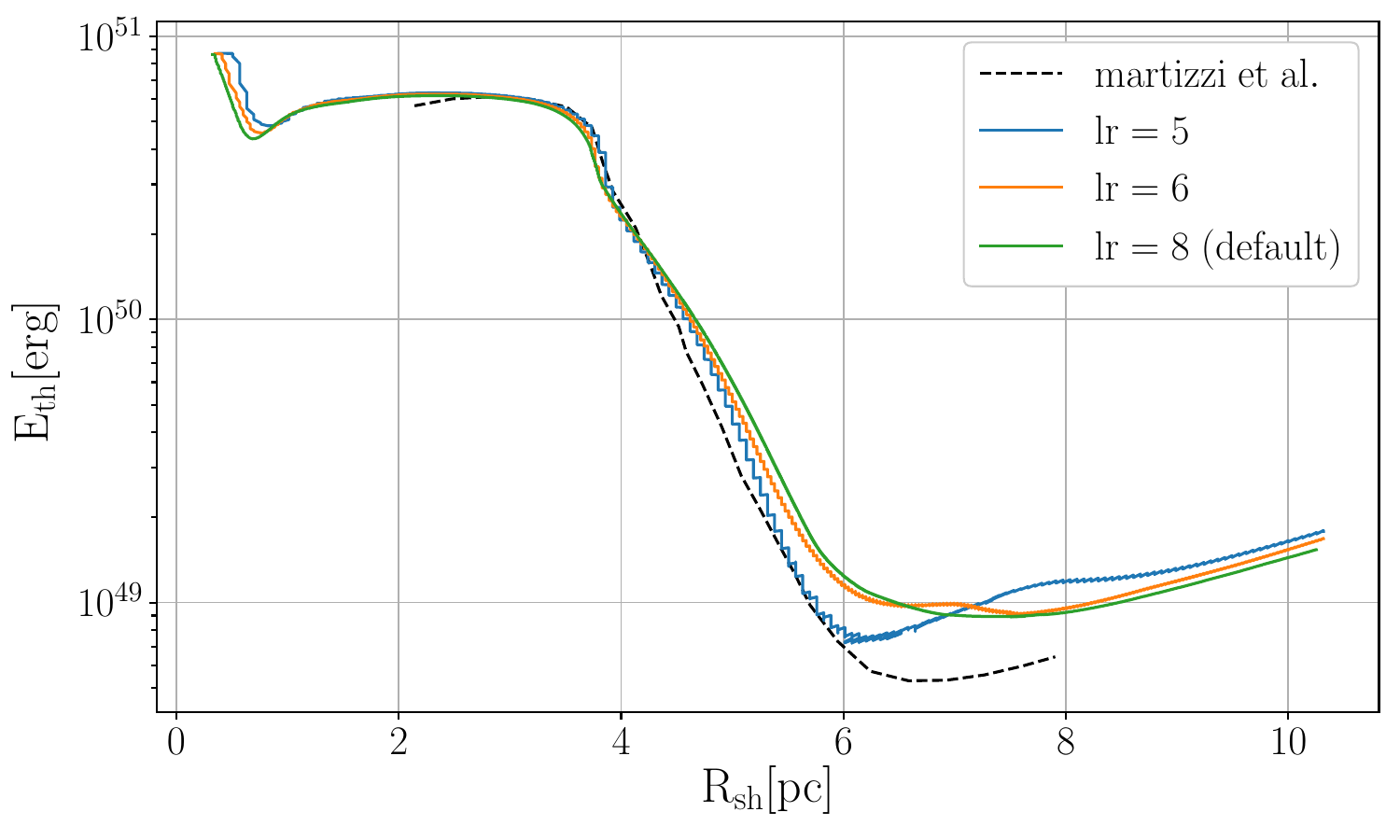}
\label{fig:res_z_eth}}%
\subfigure[Momentum]{\includegraphics[width=0.48\columnwidth]{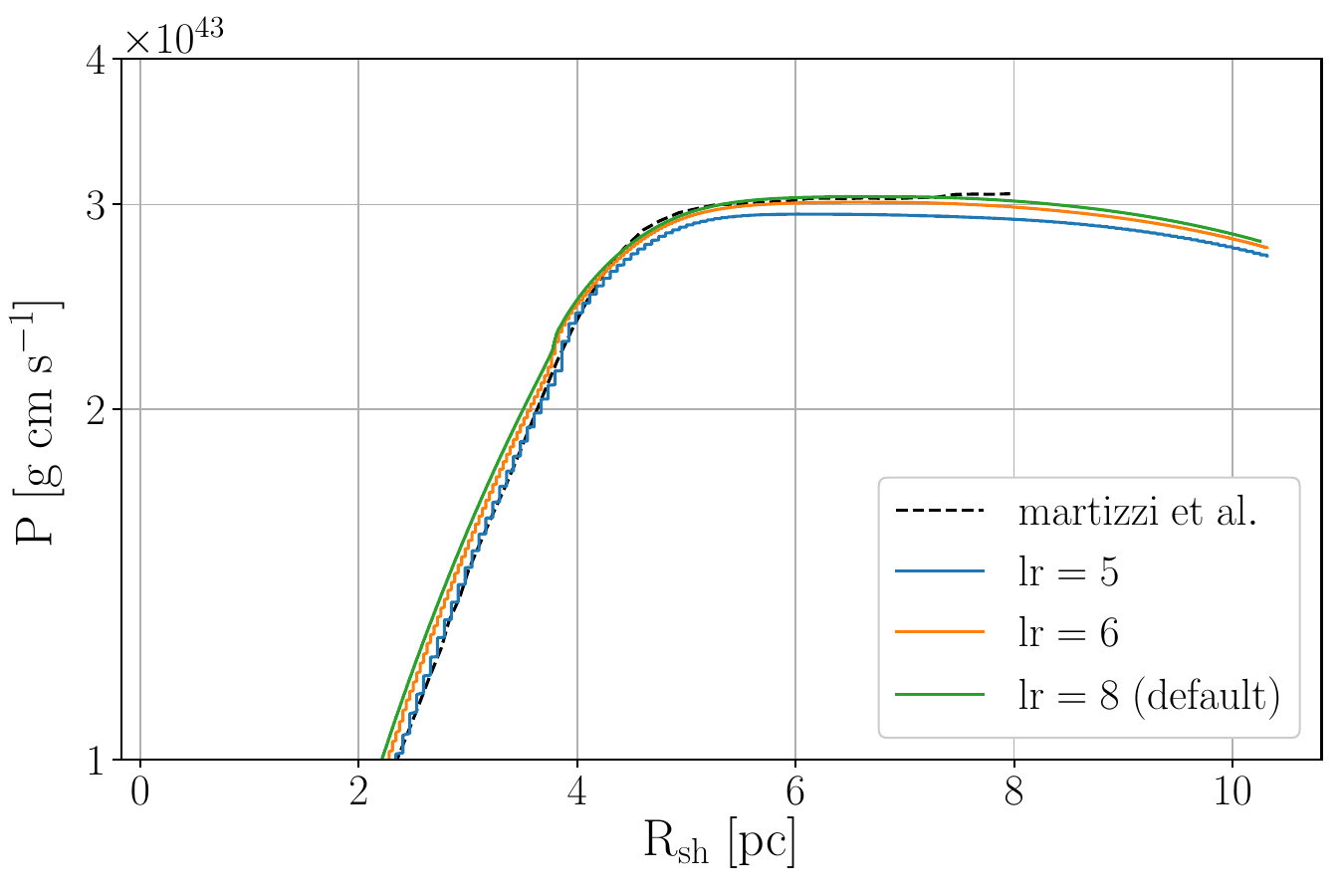}
\label{fig:res_z_momentum}}%
\caption{Thermal energy (left) and momentum (right) as a function of shock position, while varying the resolution, where $lr$ stands for \textit{levels of refinement}, starting with nbx=32 grid points. Ejecta and ISM both have $Z=Z_{\odot}$. 3D model of Martizzi et al. is also presented for comparison, plotted as a dashed line. 3D run was conducted at $512^3$ grid points, with our 1D runs at lr=5 being the closest match in resolution.}
\label{fig:res}
\end{figure*}

While we got a good match of our 1D solar-abundance ejecta exploding into solar-abundance ISM model with 3D model of \citet{MartizziSubGrid2015} for momentum, there was a significant divergence in the slopes of $E_{th}$ loss curves. Outside of us running the models in 1D, we have also ran them at a significantly higher resolution with the help of Adaptive Mesh Refinement (AMR). Thus, while the starting grid was nbx=32 points, the levels of refinement (lr) was set to 8, giving a potential resolution of 4096 grid points. \citet{MartizziSubGrid2015} presented their models at the resolution of 512 points in each direction, while not claiming absolute resolution-based convergence. In Fig.~\ref{fig:res} we present a resolution study, in which the slopes of our 1D model and the 3D model of \citet{MartizziSubGrid2015} match closely at the similar resolutions, nbx=32 with lr=5 and $512^3$ grid points respectively. Note: we reran our 1D models with \cite{SutherlandDopita1993} cooling functions instead of \cite{Gnat2012} (which is used for the results throughout the paper) to ensure consistency.

While we do not account for multidimensional effects, and the total amount of cooling is less than in 3D models, it does not affect our analysis. In this paper we primarily considered $M_{cool}$ and $R_{cool}$, which match closely in Fig.~\ref{fig:res}. In addition, for our abundance-pattern-effects study we have only been concerned with relative to solar-abundance quantities.

\vspace{-0.5cm}

\section{Abundance Spread}
\begin{figure*}[!h]
\centering
\subfigure{\includegraphics[width=0.5\columnwidth]{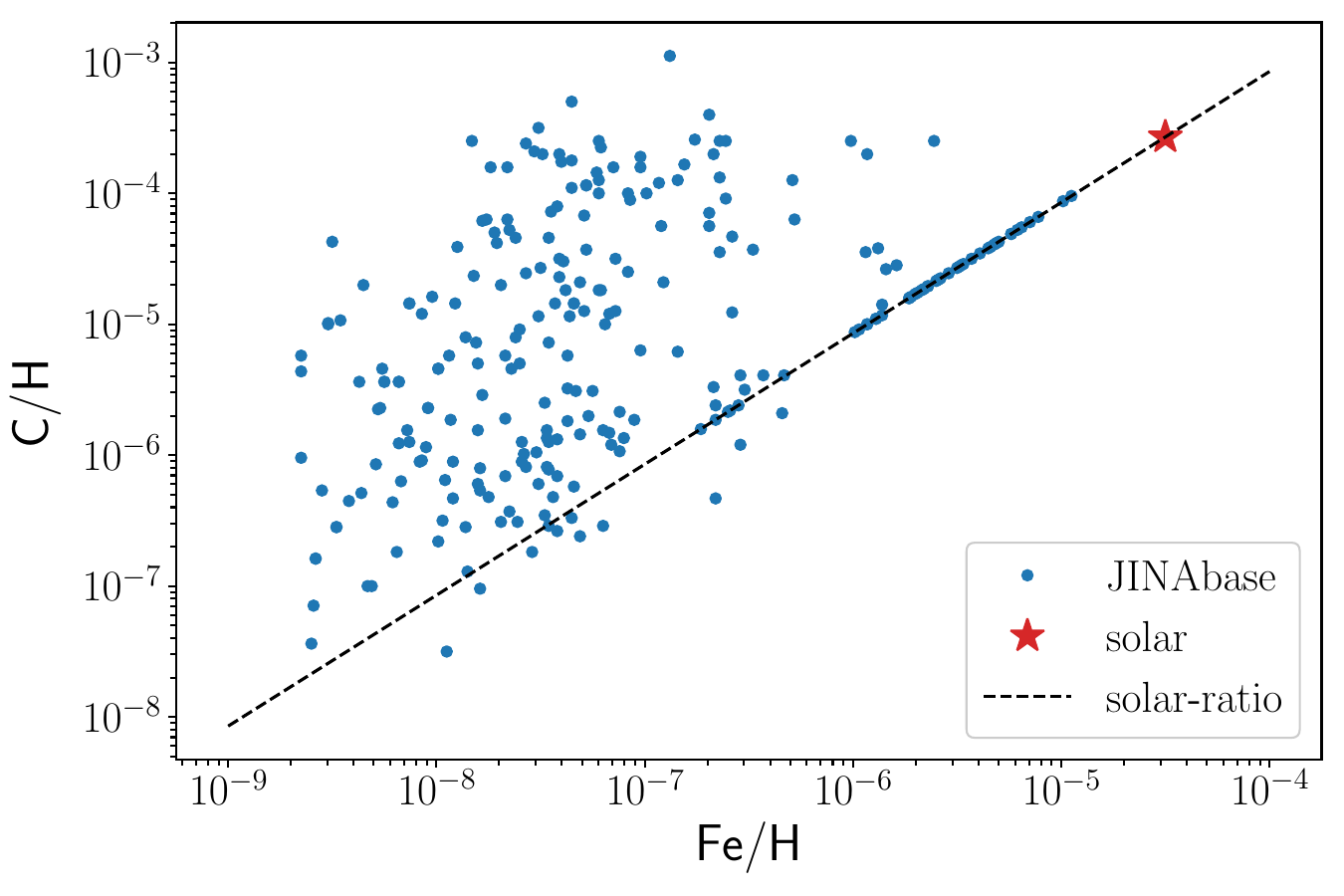}
\label{fig:C}}%
\subfigure{\includegraphics[width=0.5\columnwidth]{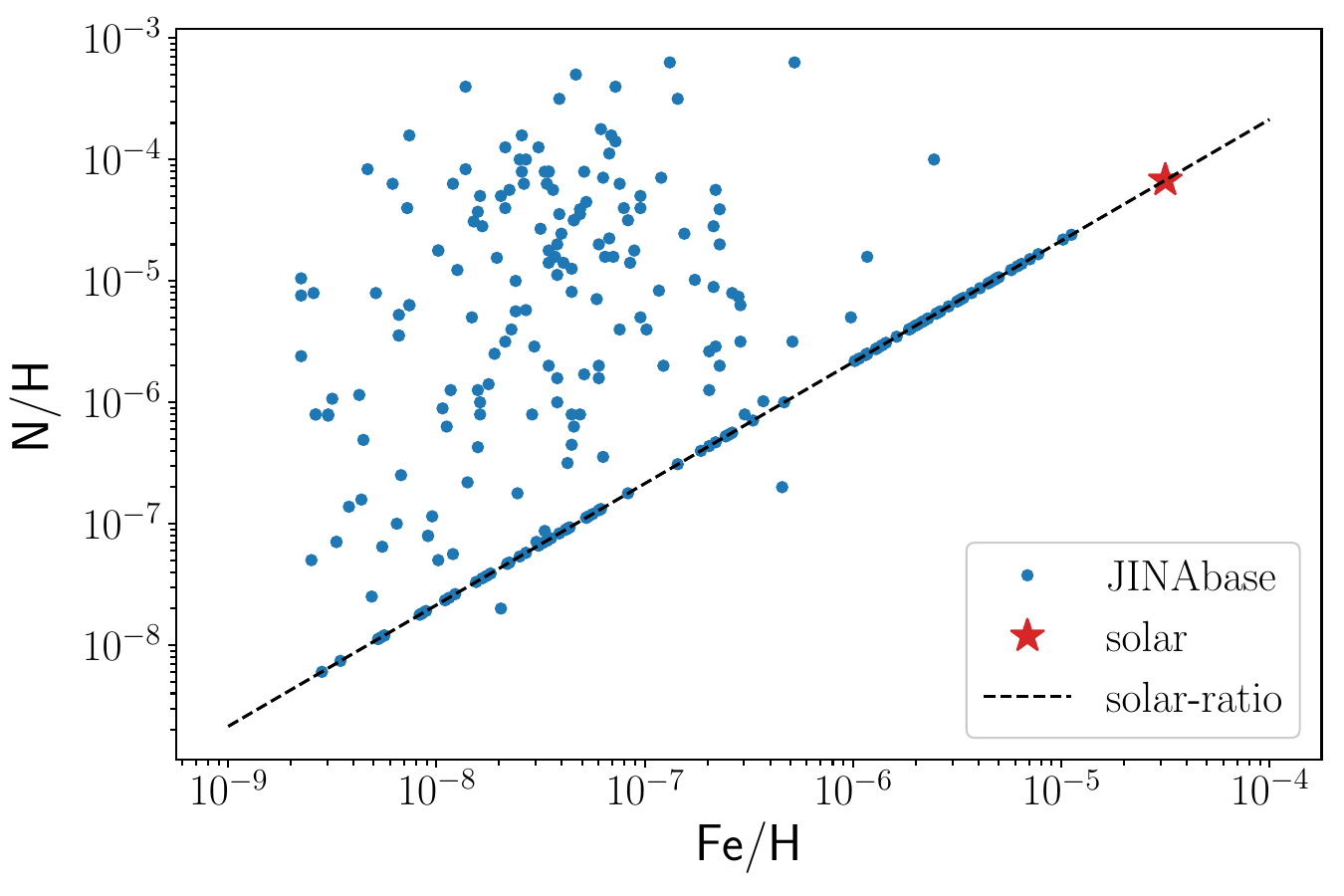}
\label{fig:N}}%
\\
\subfigure{\includegraphics[width=0.5\columnwidth]{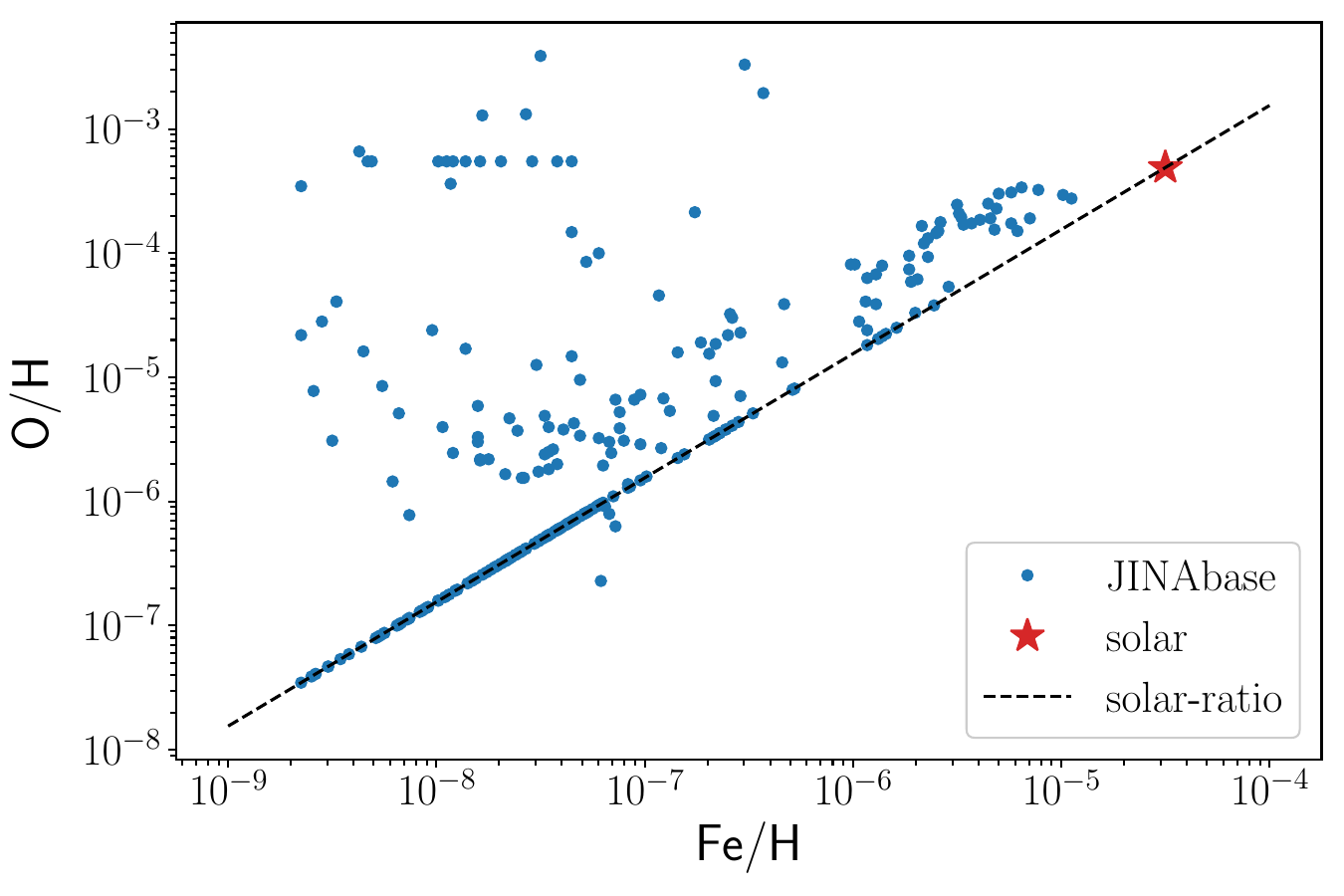}
\label{fig:O}}%
\subfigure{\includegraphics[width=0.5\columnwidth]{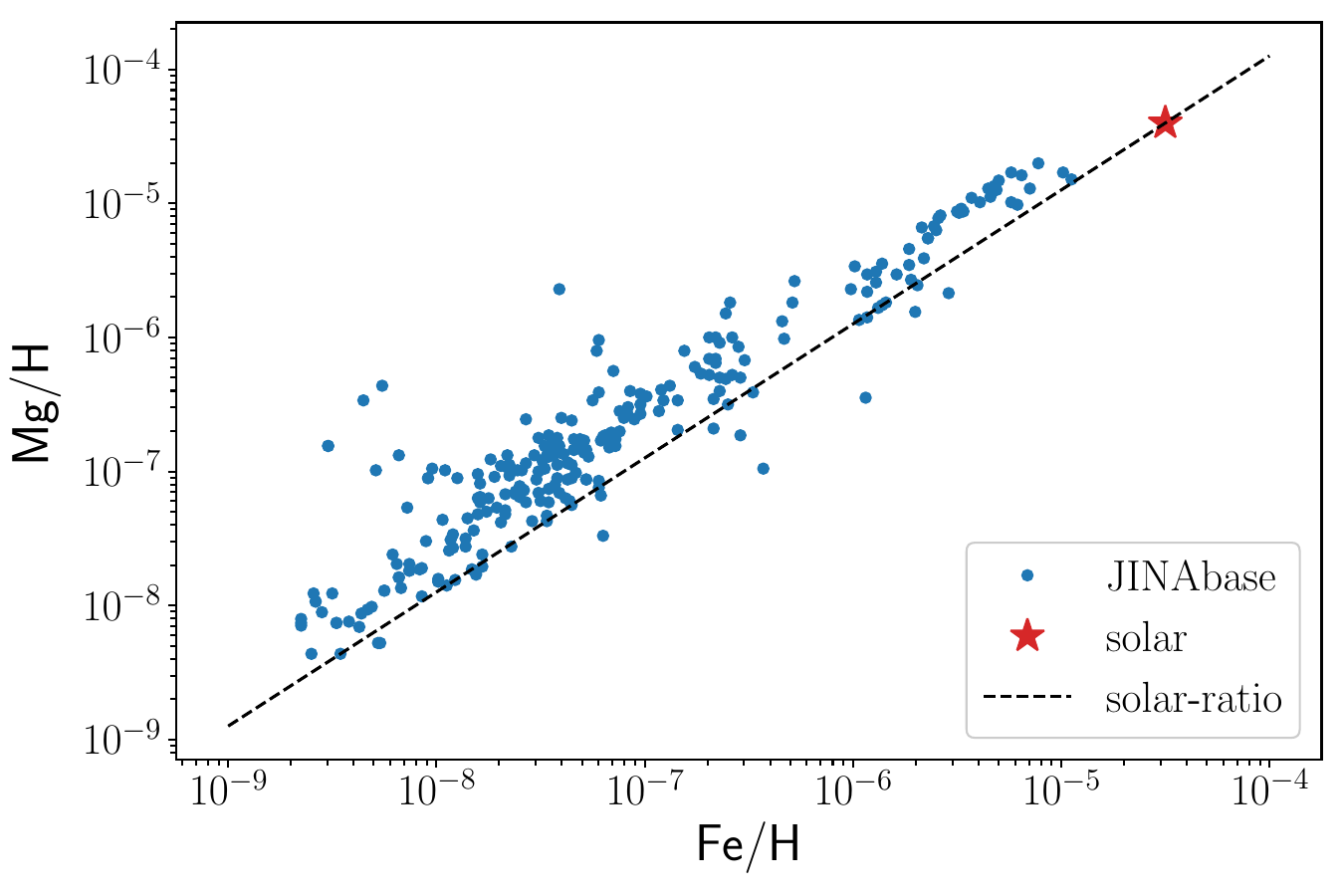}
\label{fig:Mg}}%
\\
\subfigure{\includegraphics[width=0.5\columnwidth]{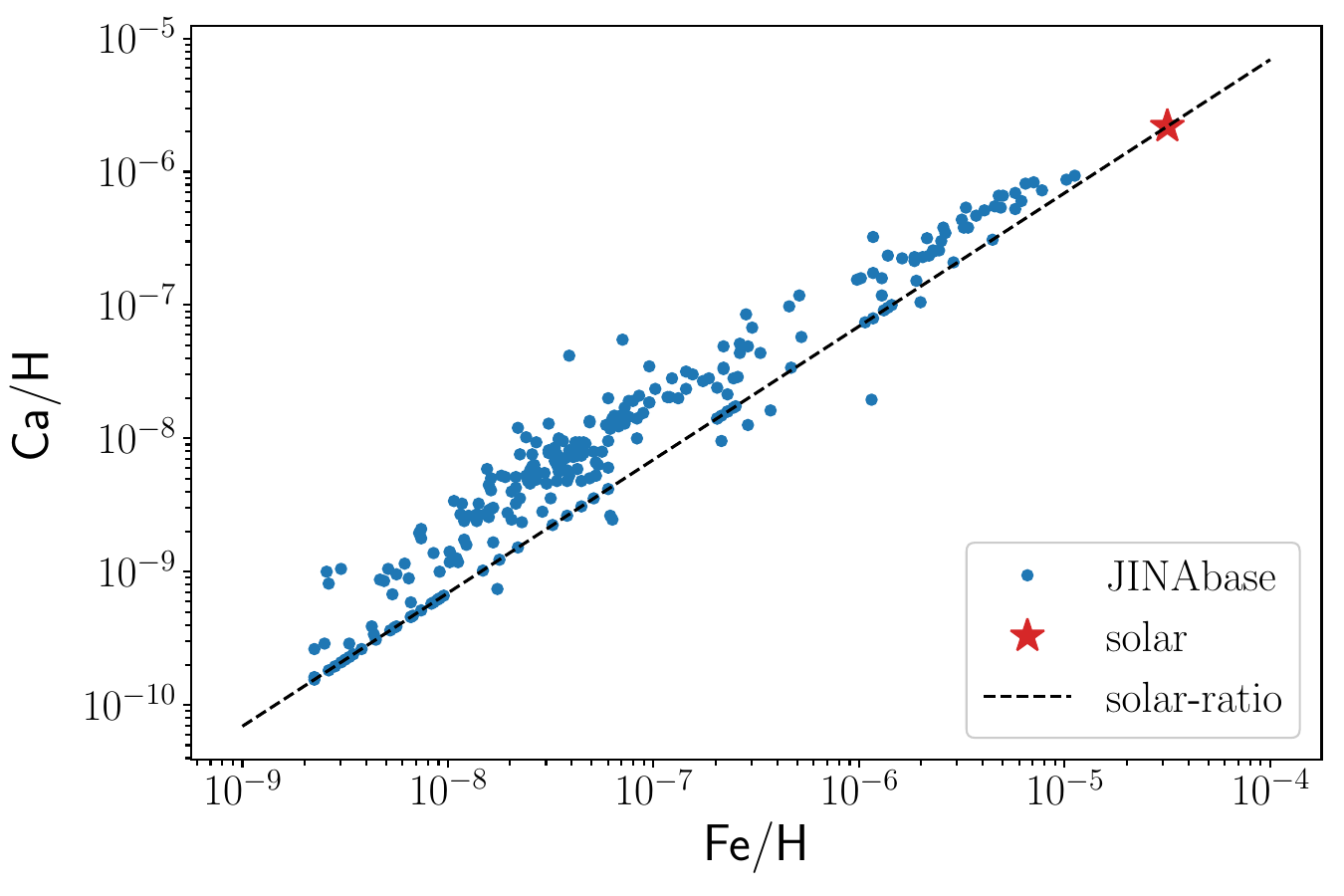}
\label{fig:Ca}}%
\subfigure{\includegraphics[width=0.5\columnwidth]{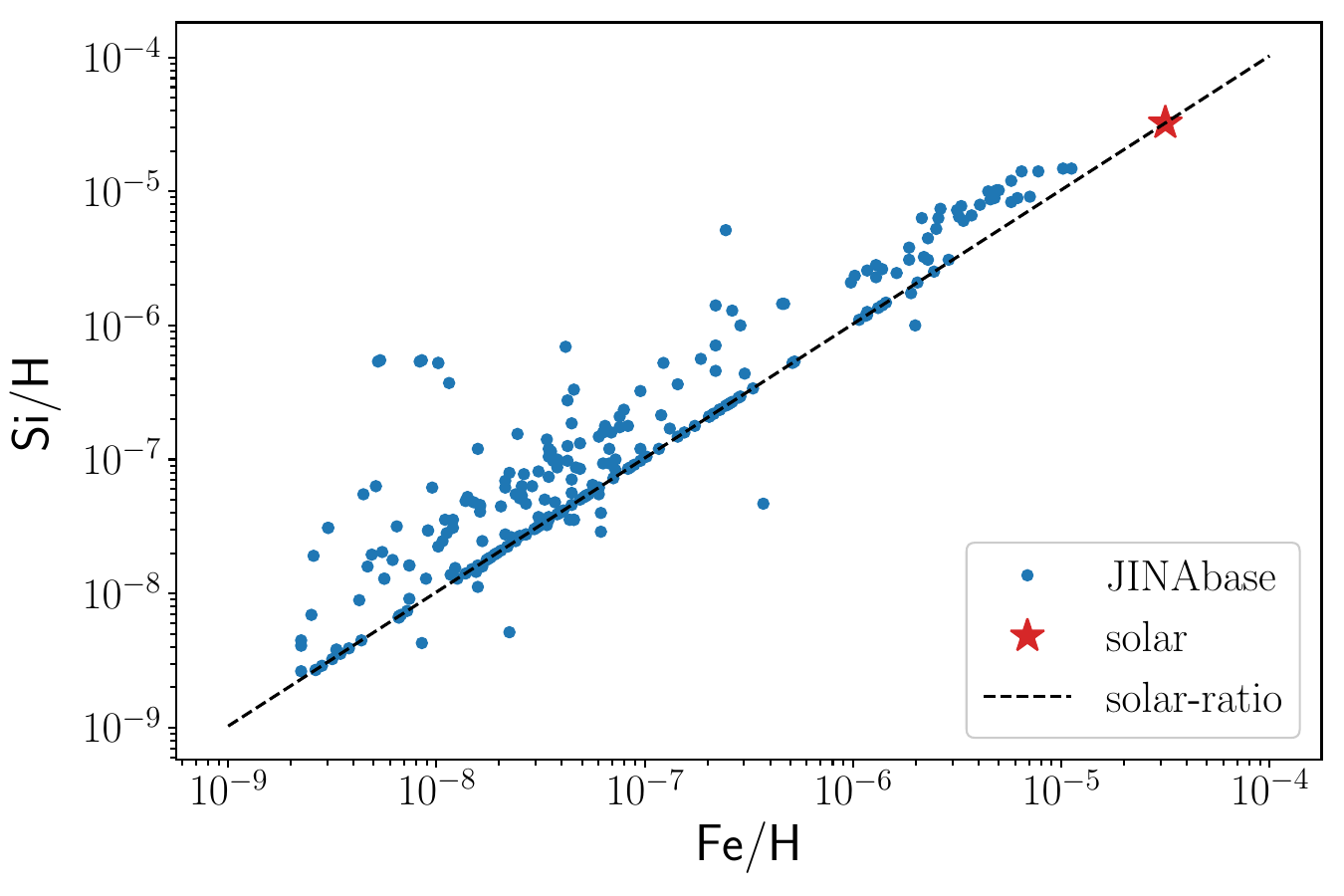}
\label{fig:Si}}%
\caption{Abundance spread of various elements with respect to Fe/H. These are selected from the stellar data from \jina\, as these stars are used as tracer particles of the primordial ISM composition for our studies. The $x$-axis representing relative Fe abundance can be thought of as a timeline for the enrichment of the Universe. It is evident that certain elements play a greater role at cooling the SNR at different times within Universe's history, than a typically used solar-pattern might suggest.}
\label{fig:abundance_spread}
\end{figure*}

The Universe has not been uniformly enriched as per solar-abundance-pattern. Looking at the low metallicity stellar data from \jina\ presented in Fig.~\ref{fig:abundance_spread} it can be seen how elemental abundance has been varying throughout time. In this case, we used stellar atmosphere measurements from the aforementioned database as tracer particles for the evolving abundance pattern of the Universe. Some elements do indeed follow the solar pattern as plotted against Fe/H, while others can experience the periods of great excess, increasing their influence on the SNR evolution. We looked for the Table \ref{tb:hierarchy} elements in \jina, and plotted a sample of abundances as an example (Ne, S, and Ar measurements are not available). The largest deviations from the solar-pattern were found to be in C, N, and O, which ultimately had the greatest effect on the variations in momentum deposition by the SNe in the low-metallicity regime.

\newpage
\vspace{-0.5cm}

\end{document}